\documentclass[a4paper,12pt,english]{article}

\usepackage{amsfonts,bm,amssymb,euscript,array,babel,cite}
\usepackage{mathtools}
\usepackage{tikz-cd}
\usepackage{amsmath}
\usepackage{tensor}
\usepackage{hhline}
\usepackage[amsmath]{empheq}
\usepackage{hyperref}
\usepackage{cancel}
\usepackage{mathrsfs}
\usepackage{pdfsync}
\usepackage{multirow}
\usepackage{tikz}

\usepackage{framed}

\newsavebox{\mysaveboxM}
\newsavebox{\mysaveboxT}

\makeatletter

\newcommand{\dd}{\mathrm{d}}
\newcommand{\DD}{\mathrm{D}}

\newcommand{\w}{\wedge}

\newcommand{\be}{\begin{equation}}
\newcommand{\ee}{\end{equation}}
\newcommand{\sfrac}[2]{{\textstyle\frac{#1}{#2}}}
\def\nn{\nonumber}
\def \bea{\begin{eqnarray}} 
\def\eea{\end{eqnarray}}
\newcommand{\mf}{\mathfrak}

\def\bi{\begin{itemize}} 
\def\ei{\end{itemize}}

\makeatother

\newtheorem{prop}[equation]{Proposition}
\newtheorem{defn}[equation]{Definition}
\newtheorem{rmk}[equation]{Remark}

\def\a{\alpha}   \def\G{\Gamma} \def\d{\delta} 
\def\e{\epsilon} 
   
\def\l{\lambda}  
    
\def\s{\sigma} \def\S{\Sigma}

 \def\Z{{\mathbb Z}} 

\def\one{\mbox{1 \kern-.59em {\rm l}}}

\numberwithin{equation}{section}

\begin{document}

\makeatother
\parindent=0cm
\renewcommand{\title}[1]{\vspace{10mm}\noindent{\Large{\bf #1}}\vspace{8mm}} \newcommand{\authors}[1]{\noindent{\large #1}\vspace{5mm}} \newcommand{\address}[1]{{\itshape #1\vspace{2mm}}}

\begin{titlepage}

\begin{flushright}
	RBI-ThPhys-2021-24
\end{flushright}

\begin{center}

\title{ {\large {Topological Field Theories induced by twisted R-Poisson  \\[4pt] structure in any dimension}}}

  \authors{\large Athanasios {Chatzistavrakidis}{\footnote{Athanasios.Chatzistavrakidis@irb.hr}} 
  }
 
 
  \address{ 
  	 Division of Theoretical Physics, Rudjer Bo\v skovi\'c Institute \\ Bijeni\v cka 54, 10000 Zagreb, Croatia \\

 }

\vskip 2cm

\begin{abstract}

We construct a class of topological field theories with Wess-Zumino term in spacetime dimensions $\ge 2$ whose target space has a geometrical structure that suitably generalizes Poisson or twisted Poisson manifolds. Assuming a field content comprising a set of scalar fields accompanied by gauge fields of degree $(1,p-1,p)$ we determine a generic Wess-Zumino topological field theory in $p+1$ dimensions with background data consisting of a Poisson 2-vector, a $(p+1)$-vector $R$ and a $(p+2)$-form $H$ satisfying a specific geometrical condition that defines a $H$-twisted $R$-Poisson structure of order $p+1$. For this class of theories we demonstrate how a target space covariant formulation can be found by means of an auxiliary connection without torsion. Furthermore, we study admissible deformations of the generic class in special spacetime dimensions and find that they exist in dimensions 2, 3 and 4. The two-dimensional deformed field theory includes the twisted Poisson sigma model, whereas in three dimensions we find a more general structure that we call bi-twisted $R$-Poisson. This extends the twisted $R$-Poisson structure of order 3 by a non-closed 3-form and gives rise to a topological field theory whose covariant formulation requires a connection with torsion and includes a twisted Poisson sigma model in three dimensions as a special case. The relation of the corresponding structures to differential graded Q-manifolds based on the degree shifted cotangent bundle $T^{\ast}[p]T^{\ast}[1]M$ is discussed, as well as the obstruction to them being QP-manifolds due to the Wess-Zumino term.   
\end{abstract}

\end{center}

\vskip 2cm

\end{titlepage}

\setcounter{footnote}{0}
\tableofcontents

\newpage

\section{Introduction}
\label{sec1}

Topological field theories play an instrumental role in a variety of physical problems in diverse spacetime dimensions. In two dimensions, the A/B-models define topological string theory \cite{Witten:1988xj,Witten:1991zz}. In three dimensions, Chern-Simons theory is directly related to general relativity \cite{Witten:1988hc}, among other applications. In four dimensions, topological Yang-Mills theory is related to the QCD theta angle and to the physics of axions. In a different direction, topological field theories can describe topological states in quantum matter, for example different types of insulators and (super)conductors in dimensions 3 to 5 \cite{Qi:2011zya,Qi:2008ew,Qi:2012cs}, and they famously serve as effective field theories capturing the response of such systems to perturbations \cite{Zhang:1988wy}.

The A-model can be alternatively viewed as the Poisson sigma model after suitable gauge fixing. The Poisson sigma model is a two-dimensional topological field theory with target space being a Poisson manifold \cite{SchallerStrobl,Ikeda}, originally introduced to account for gravity in 1+1 dimensions as a gauge theory. From a more modern point of view, it is the first instance in a tower of topological field theories of any dimension that can be described within the general framework of the AKSZ construction \cite{Schwarz:1992nx,Alexandrov:1995kv}, which was developed as a universal geometric framework of the Batalin-Vilkovisky (BV) quantization of gauge theories. Interestingly, the next instance in this tower is a three-dimensional topological field theory, the Courant sigma model, which is an essential generalization of Chern-Simons theory \cite{Ikeda:2002wh,Roytenberg:2006qz,Hofman:2002jz} with an underlying structure of an exact Courant algebroid  \cite{liu,Severa:2017oew}.  

 Such topological field theories can be extended by Wess-Zumino terms \cite{Witten:1983ar}. Typically these are topological terms supported on a spacetime of one dimension higher, whose boundary is the spacetime where the theory is defined, subject to conditions that ensure that the quantum theory is well-defined. When one introduces a Wess-Zumino term in the Poisson sigma model the resulting theory (called the $H$-twisted Poisson sigma model) is associated to a target space whose geometry departs from Poisson \cite{Klimcik:2001vg}. In other words the 2-vector structure $\Pi$ of the target space does not satisfy $[\Pi,\Pi]=0$ with respect to the Schouten-Nijenhuis bracket of multivector fields, but instead the right hand side is controlled by the closed 3-form $H$ that corresponds to the Wess-Zumino term. Such a structure is dubbed twisted Poisson in \cite{Severa:2001qm} and the corresponding manifold a twisted Poisson manifold. It is interesting to note that for this model the AKSZ construction cannot be applied directly, essentially because the 3-form obstructs the assignment of a compatible QP structure on the (graded) target space, which is one of the starting points of the method. The BV action of the $H$-twisted Poisson sigma model was found recently by a direct, traditional BV method \cite{Ikeda:2019czt}. For the Courant sigma model, a Wess-Zumino term corresponding to a closed 4-form was introduced in \cite{Hansen:2009zd}, albeit in a spirit rather orthogonal to the two-dimensional, Poisson counterpart. The underlying structure of the target space in that case was found to be that of a Courant algebroid twisted by a 4-form. Moreover, a more general approach to topological field theories with Wess-Zumino term in the framework of a generalization of the AKSZ construction was presented in \cite{Ikeda:2013wh}.

Our goal in this paper is to construct and study topological field theories with Wess-Zumino term and an underlying Poisson or twisted Poisson structure suitably embedded in a more general structure that can include higher vector fields in a dimension-independent fashion. Our motivation for this is twofold. First, we would like to answer the question: \emph{given a Poisson or twisted Poisson manifold as a target space, which topological field theories in spacetime dimension $\ge2$ exist such that their gauge symmetry is compatible with the structure on the manifold?} In other words we will be looking for topological field theories in any spacetime dimension with gauge symmetries such that their classical action functional is gauge invariant provided that the target space is equipped with \emph{some} structure that contains a Poisson or twisted Poisson 2-vector. We emphasize that once we depart from the well-known case of two dimensions, the theories we construct may naturally contain an additional higher multivector field of definite degree $p+1$ and moreover they can be equipped with a Wess-Zumino term corresponding to a $(p+2)$-form for $(p+1)$-dimensional spacetimes. This directly leads to the second motivation, since in three dimensions such a 3-vector field, usually denoted as $R$, can appear in the Courant sigma model \cite{Ikeda:2002wh,Hofman:2002jz,Roytenberg:2006qz}. From a physics standpoint, it has been argued to give rise to the so-called genuinely nongeometric $R$-flux backgrounds in string theory, see \cite{Halmagyi:2008dr,Mylonas:2012pg,Chatzistavrakidis:2015vka,Bessho:2015tkk,Heller:2016abk,Chatzistavrakidis:2018ztm} for this worldvolume perspective. Since (type II) string theory contains higher differential forms apart from the NSNS 3-form $H$, one can ask whether higher multivector fields aside the 3-vector $R$ play a similar role. Here, such $(p+1)$-vector fields will be built in the theories we will consider. 

We answer the above questions by considering a class of topological field theories in any dimension $\ge 2$ with a field content comprising suitable spacetime scalar fields, 1-forms, $(p-1)$-forms and $p$-forms. The background data that acquire a geometric interpretation in the target space of fields are a 2-vector $\Pi$, a $(p+1)$-vector $R$ and a closed $(p+2)$-form $H$ that gives rise to the Wess-Zumino term. In the general case, we will show that this class of topological field theories correspond to a target space geometry that is endowed with what we call a \emph{twisted $R$-Poisson structure}. Specifically, we define a twisted $R$-Poisson manifold $(M,\Pi,R,H)$ of order $p+1$ such that along with $\dd H=0$, $\Pi$ is a (genuine, not twisted) Poisson structure and in addition the condition 
\be \label{RPoissonintro}
[\Pi,R]=(-1)^{p+2}\langle \otimes^{p+2}\Pi,H_{p+2}\rangle\,,
\ee   
is satisfied. For vanishing $H$ and $p=2$ one obtains the models of \cite{Chatzistavrakidis:2015vka,Bessho:2015tkk,Heller:2016abk,Chatzistavrakidis:2018ztm}, which we here generalize both in presence of Wess-Zumino term and in higher dimensions. What is more, the above models suffered from the presence of a non-tensorial quantity appearing in the action functional. Here, inspired by the treatment of the twisted Poisson sigma model in \cite{Ikeda:2019czt}, we perform a detailed target space covariantization of the general class of theories we consider, including the action functional, the field equations and the gauge transformations, which requires the introduction of a connection without torsion on the target. Thus, this class of topological field theories can be thought as a generalization of the Poisson sigma model in higher dimensions.{\footnote{We note that Nambu structures is a generalization of Poisson structures in a different sense, since here we retain Poisson but generalize the topological field theory. Nambu sigma models were considered in \cite{Jurco:2012yv}.}}

One can immediately notice that twisted Poisson structures are not a special case of twisted $R$-Poisson structures for $p=1$. Since twisted Poisson sigma models exist, the question of how they fit in our approach and whether they generalize to higher dimensions naturally arises. We answer this by exploring deformations of the general topological field theories we constructed. It is shown that there exists only a small number of such deformations with our choice of field content and moreover they only appear in special number of dimensions, namely 2, 3 and 4. We perform a detailed analysis of the corresponding ``islands'' of theories in these dimensions. In particular, in two dimensions the twisted Poisson sigma model is recovered, along with a cousin theory that involves two independent 2-vector fields on the target and corresponds to a twisted $R$-Poisson manifold of order 2. In three dimensions, we find that apart from the already covered case of twisted $R$-Poisson structure of order 3, which involves a Poisson 2-vector, there exists an extension to a theory with a twisted Poisson 2-vector. This is twisted by a 3-form, which in this case is \emph{not} the Wess-Zumino term, since the latter is a 4-form. We call the corresponding structure a bi-twisted $R$-Poisson one, since both $\Pi$ and $R$ are twisted in a certain sense by different degree differential forms. Specifically, the structure is $(M,\Pi,R,S,H)$, where $S$ is a 3-form such that $\dd S=-H$ and 
\bea \label{bitRpoissonintro}
\frac 12 \, [\Pi,\Pi]&=&R+\langle \Pi\otimes \Pi\otimes \Pi,S\rangle\,.
\eea 
One can now immediately see that for $H=0$ and $R=0$, one obtains a twisted Poisson structure in the ordinary sense, albeit in a three-dimensional topological field theory. Thus, the three-dimensional case contains 
 examples of theories with an underlying (i) twisted Poisson or (ii) twisted $R$-Poisson, or (iii)  bi-twisted $R$-Poisson structure on the target space. In the last two cases, a central role is played by the Wess-Zumino term, thus realizing and extending the approach of \cite{Hansen:2009zd}. In addition, we discuss the target space covariant formulation of these theories, in particular for the bi-twisted case where a connection with torsion is necessary.   
       
We accompany our analysis with discussions of the structures we uncover in terms of graded geometry. Specifically, we describe (bi-)twisted $R$-Poisson manifolds as differential graded supermanifolds, namely as $Q$-manifolds, and determine the (co)homological vector field that defines them. The graded target space turns out to be the degree shifted cotangent bundle $T^{\ast}[p]T^{\ast}[1]M$ for any $p\ge 1$. Being cotangent bundles, such target spaces naturally carry a symplectic (P-)structure too. However, in presence of the $(p+2)$-form $H$, the two structures are not compatible and the graded manifold fails to be symplectic, i.e. a QP-manifold, and it becomes one only when $H$ vanishes. In the latter case, one may employ the AKSZ construction to determine the BV action of the models. However, this is not the case in general, for the same reason as for the twisted Poisson sigma model in two dimensions \cite{Ikeda:2019czt}. We nod at the construction of the BV action in the conclusions section, where we mention the challenges and the differences to the more tractable two-dimensional case.  

The rest of the paper is organized as follows. In Section \ref{sec2} we introduce the general case of topological field theories with twisted $R$-Poisson structure. First, we start with a brief recapitulation of the twisted Poisson sigma model in two dimensions in Section \ref{sec21}, highlighting the features that will be kept or lost in higher dimensions. Section \ref{sec22} contains the topological field theories induced by twisted $R$-Poisson structures in dimensions $p+1\ge 2$ in a non manifestly target space covariant formulation that requires to restrict on a local patch, along with their gauge symmetries and field equations. In Section \ref{sec3} we covariantize the gauge transformations, field equations and the action functional of the theories by means of a connection without torsion on the target space, thus providing a solid basis of the models beyond local coordinate patches. Section \ref{sec4} contains a discussion of the target space as a $Q$-manifold and its P-structure. In Section \ref{sec5} we discuss special cases and examples. In particular, we first classify the possible deformations that are allowed by the field content of the theories we study in special spacetime dimensions (Section \ref{sec51}). In Section \ref{sec52} we discuss examples in 2, 3 and 4 dimensions. In the two-dimensional case we emphasize how the distinct structures of twisted Poisson and twisted $R$-Poisson come together in our approach. In the three-dimensional case, we perform a more in-depth analysis, since there are three distinct structures that can be realized, namely twisted $R$-Poisson, twisted Poisson and bi-twisted $R$-Poisson. Since the latter is the most general one, we present it in detail and find its target space covariant formulation and its $Q$-structure. In four dimensions, we mainly discuss the twisted $R$-Poisson case and briefly comment on the single admissible deformation and its consequences. Finally, Section \ref{sec6} contains our conclusions and outlook to further work.

\section{TFTs with WZ term \& twisted R-Poisson structure} 
\label{sec2}

\subsection{The Wess-Zumino Poisson Sigma Model in 2D}
\label{sec21} 

We begin our analysis with the $H$-twisted Poisson Sigma Model (HPSM), which is a topological field theory with Wess-Zumino term in two dimensions and a target space equipped with a twisted Poisson structure. Although none of the material regarding the HPSM is new, in the discussion that follows we offer a perspective to the model which is close to the spirit of the higher-dimensional topological field theories discussed in the rest of this paper. Moreover, this discussion will help us introduce some notation and conventions useful for the rest of the analysis. 

The field content of the HPSM comprises a set of scalar fields $X^{i}$, the components of a map $X:\S_2\to M$ from a two-dimensional spacetime $\S_2$ to a smooth manifold $M$, the target space, and a set of spacetime 1-forms $A\in \Omega^{1}(\S_2,X^{\ast}T^{\ast}M)$ with values in the pull-back bundle of the cotangent bundle of $M$ by $X$. It will often prove useful to introduce local coordinates, respectively local bases, for manifolds and vector bundles that will be encountered, although in the end we will always present the basis-independent results. To this end, we introduce local coordinates $(\s^{a}), \a=0,1$ for $\S_2$ and $(x^{i}), i=1,\dots,\text{dim}\,M$ for $M$. The spacetime scalars $X^{i}=X^{i}(\s^{\a})$ are simply the pull-backs of the latter by the map $X$, i.e. $X^{i}=X^{\ast}(x^{i})$, as usual. Regarding the 1-forms, we introduce a local basis $e^{i}$ of the pull-back bundle $X^{\ast}T^{\ast}M$ given some basis $\mf{e}^{i}$ of $T^{\ast}M$, in which case we can write $A=A_i\otimes e^{i}$. Note that $A_i$ are spacetime 1-forms and therefore may be expanded in turn as $A_i=A_{i\a}(\s)\dd\s^{\a}$, although we will mostly be working already in a coordinate-independent formalism with respect to spacetime.  

With the above field content, one may write down a two-dimensional topological field theory with classical action functional 
\be \label{SHPSM}
S_{\text{HPSM}}=\int_{\S_{2}}\left(A_i\w\dd X^{i}+\frac 12 \, (X^{\ast}\Pi^{ij})A_i\w A_j\right)+\int_{\S_3}X^{\ast}H\,,
\ee  
where $\dd$ is the two-dimensional exterior derivative and $\S_3$ is an open membrane whose boundary is $\S_2$. Clearly the background field $X^{\ast}\Pi^{ij}\equiv \Pi^{ij}(X)$ is antisymmetric and therefore it geometrically corresponds to the pull-back by the map $X$ of the components of a 2-vector $\Pi\in \bigwedge^2(TM)$. We avoid introducing different notation for the 2-vector and its pull-back, and similarly for other structures to be encountered below, to avoid clutter; the nature of each quantity of that sort should be clear from context and in any case we will be mostly referring to pull-back objects unless otherwise stated. The second ingredient is the Wess-Zumino term, supported on the membrane $\S_3$. This is the pull-back of a closed 3-form $H$ on $M$, hence $\dd H=0$, and does not depend on the membrane as long as $H$ defines an integer cohomology class, in which case the above action makes sense as a two-dimensional topological field theory and its path integral is not ambiguous \cite{Witten:1983ar}. 

The target space of the theory is a twisted Poisson manifold $(M,\Pi,H)$, namely $M$ is equipped with an antisymmetric 2-vector $\Pi$ and a 3-form $H$ such that 
\be \label{twistedPoisson}
\frac 12 \, [\Pi,\Pi]=\langle \otimes^{3}\Pi,H\rangle, 
\ee  
where the bracket on the left hand side is the Schouten-Nijenhuis bracket of multivector fields and
 the contractions among the three appearances of the 2-vector and the 3-form on the right hand side are in the odd order of indices in local coordinates, here the first, third and fifth index in $\otimes^{3}\Pi$. The angle brackets denote the canonical inner product between sections of the tangent and cotangent bundles of $M$. The field theory given by \eqref{SHPSM} arises as the one which is invariant under a gauge symmetry that corresponds to $\S_2$-dependent deformations along the leaves of the foliation of $M$ generated by the above twisted Poisson structure. In other words, including local transformations of the 1-forms, $S_{\text{HPSM}}$ is a gauge invariant functional under the following set of gauge transformations
\bea 
\d X^{i}&=&\Pi^{ji}\epsilon_j\,,\\ \label{HPSMtrafo2}
\d A_i&=& \dd\epsilon_i+\partial_i\Pi^{jk}A_j\epsilon_k+\frac 12\, \Pi^{jk}H_{ijl}(\dd X^{l}-\Pi^{lm}A_m)\epsilon_k\,,
\eea   
where $\epsilon_i=\epsilon_i(\s^{\a})$ is the $\S_2$-dependent scalar gauge parameter. We note that for $H=0$ this results in the ordinary Poisson sigma model, where $\Pi$ is a Poisson structure and the transformation of the 1-form $A_i$ is akin to ordinary nonlinear gauge theory.  

Having introduced the HPSM 
we briefly discuss two further issues that will be important in later sections. First, note that the classical field equations stemming from the action functional $S_{\text{HPSM}}$  as Euler-Lagrange equations are 
\bea 
F^{i}&:=&\dd X^{i}+\Pi^{ij}A_j=0\,,
\\ \label{HPSMeom2}
G_i&:=&\dd A_i+\frac 12 \, \partial_i\Pi^{jk}A_j\w A_k+\frac 12 \, H_{ijk}\dd X^{j}\w\dd X^{k}=0\,.
\eea  
In physics terminology, we refer to $F^{i}$ and $G_{i}$ as the field strengths of $X^{i}$ and $A_i$ respectively. 
It is then simple to confirm that due to the basic structural equation \eqref{twistedPoisson}, both field equations transform covariantly under gauge transformations. With the above definition of the field strength $F^{i}$ of $X^{i}$, which may be written in a basis-independent way as $F=\dd X+\Pi(\cdot,A)$,  it is obvious that $S_{\text{HPSM}}$ may be rewritten as 
\be \label{SHPSMalt}
S_{\text{HPSM}}=\int_{\S_{2}}\left(A_i\w F^{i}-\frac 12 \, \Pi^{ij}(X)A_i\w A_j\right)+\int_{\S_3}X^{\ast}H\,.
\ee 
Although this rewriting of the HPSM action in terms of the lowest degree field strength looks like a triviality, a similar, non-trivial rewriting will turn out to be instrumental in establishing target space covariance for topological field theories having an underlying Poisson structure in general dimensions. For the case at hand, this rewriting is of course not important and we state it here as a simple observation to be appreciated later on. 

Second, one would like to have a theory in terms of basis-independent tensorial quantities in target space. For the action itself, this is simple in the present case \cite{Ikeda:2019czt}. Indeed, first recall that $\dd X$ is a linear map from $T_P\S_2$ to $T_{X(P)}M$ for every $P\in \S_2$ and therefore the spacetime 2-form $A\w \dd X$ is well defined through 
\be 
(A\w \dd X)(v_1,v_2)=A(v_1)(\dd X(v_2))-A(v_2)(\dd X(v_1))\,,
\ee   
for vector fields $v_1,v_2$ in $T\S_2$ \cite{Fulp:2002fm}.  Then the inherently target space covariant form of the HPSM action functional is \cite{Ikeda:2019czt}
\be 
S_{\text{HPSM}}=\int_{\S_2}\left(\langle A,\dd X\rangle + \frac 12 \, (\Pi\circ X)(A,A)\right)+\int_{\S_3}X^{\ast}H\,,
\ee 
where the angles denote the natural inner product of the pullback bundles associated to $TM$ and $T^{\ast}M$, and the composition $\Pi\circ X$ gives $(\Pi\circ X)(A,A)= (X^{\ast}\Pi^{ij})A_i\w A_j$. As discussed in \cite{Fulp:2002fm}, one can define the spacetime 2-form $\Pi_{X}(A\w A)$ through 
\be 
\Pi_{X}(A\w A)(v_1,v_2)=(\Pi\circ X)(A(v_1),A(v_2))\,
\ee 
in which case $(\Pi\circ X)(A\w A)=\frac 12 \Pi_{X}^{ij}A_i\w A_j$. Hereby we follow the former notation.
We also note in advance that the simplicity of passing to the target space covariant form of the action in the present case is lost in the more general topological field theories to be discussed below and one should reside in the alternative expressions that correspond to \eqref{SHPSMalt} in those cases, as explained in detail in the ensuing. 

Besides the action, one should also address target space covariance for expressions like the field equation \eqref{HPSMeom2} and the gauge transformation \eqref{HPSMtrafo2} of the 1-form. Here we simply state the corresponding results from \cite{Ikeda:2019czt} without providing many details, since we will have to address the same issues for the topological field theories constructed below where we will be more thorough. To this end, Ikeda and Strobl \cite{Ikeda:2019czt} introduce a connection $\nabla$ on $TM$, which has coefficients $\G_{ij}^{k}$ in a holonomic frame, that is $\nabla\partial_i=\G_{ij}^{k}\dd x^{j}\otimes \partial_k$---recall that $(x^{i})$ are local coordinates on $M$. In the twisted case, this is a connection with torsion, the latter being controlled by the 3-form $H$, in particular $\G_{ij}^k=\mathring{\G}_{ij}^{k}-\sfrac 12 \Pi^{kl}H_{ijl}$ with $\mathring{\G}_{ij}^{k}$ the coefficients of an arbitrary connection $\mathring{\nabla}$ without torsion. Moreover, it naturally induces a connection on $T^{\ast}M$ and in the holonomic frame $\nabla\dd x^{i}=-\G^{i}_{jk}\dd x^j\otimes \dd x^{k}$. Then, denoting by $\d^{\nabla}$ the corresponding transformation with the auxiliary connection $\nabla$, it turns out that the gauge transformation of the $X^{\ast}T^{\ast}M$-valued 1-form $A$ is 
\be \label{covtrafoAHPSM}
\d^{\nabla}A=\DD \epsilon-(T\circ X)(A,\epsilon)\,,
\ee    
where $\DD$ is the exterior covariant derivative on differential forms induced by $\nabla$ and $T$ is the torsion of the $T^{\ast}M$-covariant derivative $^{T^{\ast}M}\nabla_{e}e':=\nabla_{\Pi(e)}e'$, with $e,e'\in \Omega^{1}(M)$. This vector bundle torsion, not to be confused with the torsion of $\nabla$ that was mentioned before, turns out to be independent of $H$ and specifically it is the opposite of the covariant derivative of the twisted Poisson 2-vector with respect to $\mathring{\nabla}$, namely $T=-\mathring{\nabla}\Pi$. Thus, note that $H$ resides only in the first term of the right-hand side in \eqref{covtrafoAHPSM}. Finally, with regard to the field equation for $A$, one should define a target space covariant field strength. The connection introduced above serves this purpose too. Indeed, the general form of the covariant field strength is found to be 
\be 
G=\DD A-\frac 12 (T\circ X)(A,A)\,. 
\ee   
We will revisit these statements in a more general context in Section \ref{sec22}. Note that the above obviously continue to hold in case $H=0$, namely for the (untwisted) PSM, in which case $\Pi$ is an ordinary Poisson 2-vector and $\nabla=\mathring{\nabla}$. 
 
\subsection{WZ-TFTs with twisted R-Poisson structure in any dimension} 
\label{sec22}

Our purpose now is to examine what is in a certain sense the analogue of the (H)PSM in dimensions greater than 2. In other words, we are looking for topological field theories defined on $(p+1)$-dimensional spacetime $\S_{p+1}$, corresponding to sigma models associated to a map $X:\S_{p+1}\to M$ from a $p$-brane to a target space with suitable structure that includes a Poisson or twisted Poisson 2-vector. The (H)PSM in two dimensions is a special model where $p=1$, essentially $p$ denoting the degree of the highest form in the field content of the theory.{\footnote{To avoid confusion, the class of Wess-Zumino twisted field theories we consider in this section will not contain the HPSM as a special case per se, rather they are inspired by it. The HPSM will be included through the special class of deformed models described in Section \ref{sec4}. Nevertheless, although the $p=1$ case does not result in the HPSM, it yields a theory of two target space 2-vectors coupled in a specific way. We discuss this example in Section \ref{sec4} and clarify there how the HPSM becomes a member of this class of theories upon deformation.}} For the theories we study, we consider a Wess-Zumino term on a $(p+2)$-brane whose boundary is $\S_{p+1}$. Naturally, this corresponds to a closed $(p+2)$-form $H_{p+2}$, such that $\dd H_{p+2}=0$. In the following we will simply denote this as $H$ without a subscript, its degree being obvious from the context. One of the main questions we address first is which structural identity replaces the twisted Poisson one, namely \eqref{twistedPoisson}, and consequently what is the structure of the target space $M$.    

Let us begin with the field content of the theories we are going to construct. First, this includes the usual spacetime scalar fields corresponding to the components $(X^{i})$ of the sigma model map $X:\S_{p+1}\to M$, being $\text{dim}\,M$ in number. In addition we consider the following three sets of spacetime differential forms:
\begin{itemize} 
	\item $A\in \Omega^{1}(\S_{p+1},X^{\ast}T^{\ast}M)$. These are 1-forms in spacetime $\S_{p+1}$ that take values in the pull-back of the cotangent bundle $T^{\ast}M$ by $X$. Introducing local coordinates $(\s^{\a}), \a=0,\dots,p$ on $\S_{p+1}$ and a basis $e^{i}$ on $X^{\ast}T^{\ast}M$, we may write $A=A_{i\a}(\s)\, \dd\s^{\a}\otimes e^{i}$. 
	\item $Y\in \Omega^{p-1}(\S_{p+1},X^{\ast}TM)$. These are $(p-1)$-forms in $\S_{p+1}$ taking values in the pull-back of the tangent bundle $TM$ by $X$, thus $Y=Y^{i}_{\a_1\dots\a_{p-1}}(\s)\dd\s^{\a_1}\w\dots\w\dd\s^{\a_{p-1}}\otimes e_i$, where $e_i$ is the basis of $X^{\ast}TM$ dual to $e^{i}$.  
	\item $Z\in \Omega^{p}(\S_{p+1},X^{\ast}T^{\ast}M)$. There are $p$-forms in $\S_{p+1}$ with values in the pullback of $T^{\ast}M$, like $A$. 
	\end{itemize} 
In the rest of this section we work directly in a coordinate-independent formalism for spacetime, albeit in local bases for the target space and its (co)tangent bundles. Therefore, initially we study the formulation of the theories in a local patch of $M$ and leave the study of the global structure for Section \ref{sec3}. Thus the field content we consider is 
$(X^{i},A_{i},Y^{i},Z_{i})$ 
of form degrees $(0,1,p-1,p)$ respectively.{\footnote{{Certainly this brings to mind the coordinate structure of a class of Q(P) manifolds; indeed, we will discuss this aspect in Section \ref{sec3} too.}} 
	
One may directly construct a general candidate topological field theory with the above field content simply by combining the form degrees of the fields above and identifying the admissible terms. These are terms that exist in any dimension of $\S$; certainly it can happen that for special number of dimensions there exist additional possibilities. We refer to these extra admissible terms as deformations and classify them for the given field content in Section \ref{sec4}. In the general case (without additional deformations), the corresponding classical action functional for $p>0$ is 
\bea 
S^{(p+1)}&=& \int_{\S_{p+1}}\left(Z_i\w\dd X^{i}-A_i\w \dd Y^{i}+\Pi^{ij}(X)\, Z_i\w A_j+\frac 12 \, Q^{ij}_{k}(X)\, Y^{k}\w A_i\w A_j\, + \right. \nn \\[4pt] && \quad \qquad \left. +\, \frac 1{(p+1)!}R^{i_1\dots i_{p+1}}(X)\, A_{i_1}\w\dots \wedge A_{i_{p+1}}\right)+\int_{\S_{p+2}}X^{\ast}H\,. \label{Sp+1}
\eea 
As before, $\dd$ is the exterior derivative on $\S_{p+1}$ and $H$ is a closed $(p+2)$-form on $M$. The closedness of the $(p+2)$-form ensures that the variation of the Wess-Zumino term drops to the boundary and hence it contributes to the field equations through the map $X$ and not through its extension that defines the higher-dimensional term in the action. Moreover, as in the HPSM, there are additional conditions that ensure that the quantum theory of the model is well-defined \cite{Figueroa-OFarrill:2005vws}. Existence of the extension to $\S_{p+2}$ that defines the term requires that the homology class $[X(\S_{p+1})]\in H_{p+1}(M)$ vanishes. Independence on the choice of $\S_3$ requires that $H$ defines an integer cohomology class, specifically $[H]/2\pi\in H^{{p+2}}(M,\Z)$. 

In the action functional \eqref{Sp+1} we also encounter three $X$-dependent coefficients, one for each of the possible terms that involve the differential forms of the field content.  At this stage we make the following assumptions. First, $R^{i_1\dots i_{p+1}}$ evidently corresponds to the pullback of the components of an antisymmetric $(p+1)$-vector $R\in \G(\bigwedge^{p+1}TM)$ on $M$. The Schouten-Nijenhuis bracket $[R,R]$ is an antisymmetric $(2p+1)$-vector on $M$. Certainly, if $\text{dim}\,M < 2p+1$ then $[R,R]=0$, a condition referred to as generalised Poisson in \cite{deAzcarraga:1996qsd}. However, for $\text{dim}\,M\ge 2p+1$ we note that it is not a necessary condition for what follows, since the Schouten-Nijenhuis bracket of $R$ with itself will not appear in the analysis and hence we may allow it to be an arbitrary $(2p+1)$-vector. 
Furthermore, with regard to $S^{(p+1)}$ we assume that $\Pi^{ij}$ are the components of a (untwisted) Poisson 2-vector and therefore satisfy $[\Pi,\Pi]=0$, a condition that can be eventually relaxed only in special cases (including $p=1$, as expected, but also $p=2$.)

Finally, the remaining coefficients are taken to be $Q_{k}^{ij}=-\partial_k\Pi^{ij}$, namely the structure functions appearing already in the transformation of the gauge field $A_i$ of the HPSM in its non-covariant formulation. One can immediately complain that with this choice we have introduced a non-tensorial object in the action functional and that $S^{(p+1)}$ makes sense only in a coordinate patch but not in general. However, we will prove in Section \ref{sec3} that this is not the case and that when subtle points regarding target space covariance are taken care of, this action makes sense and a manifestly covariant formulation for it follows. Note that on the one hand these are precisely the same issues one encounters for the (H)PSM, where as we already discussed the field equations and the transformation rules obtained at face value in its original formulation are not manifestly covariant. On the other hand, the situation in the present case is even more complicated, since non manifestly covariant terms appear already at the level of the action. All these issues are resolved by introducing an auxiliary connection in the target space as we will discuss extensively in Section \ref{sec3}. 

What remains unspecified is the Schouten-Nijenhuis bracket $[\Pi,R]$, which is an antisymmetric $(p+2)$-vector. Indeed, this is the main interesting aspect that the non manifestly covariant formulation directly uncovers. Studying the gauge symmetries of the model, we will see that the structural condition relating the three structures $\Pi, R$ and $H$ is \eqref{RPoissonintro}, which we reproduce here:
\be \label{RPoisson}
 [\Pi,R]=(-1)^{p+2}\langle \otimes^{p+2}\Pi,H_{p+2}\rangle\,,
\ee   
where for this crucial equation we reinstated the degree of $H$ for clarity and the contractions of the $p+2$ copies of $\Pi$ with the $(p+2)$-form on the right hand side are over the odd-order factors of the even degree $(2p+4)$-tensor $\otimes^{p+2}\Pi$. (If one wishes to get rid of the sign on the right hand side, the opposite convention can be used, but we avoid this to be aligned with the convention used already in the case of twisted Poisson structures.) To summarize the structure of the target space, we employ the following definition.

\begin{defn}\label{tRPoisson}
	A twisted $R$-Poisson manifold of order $p+1$ is a quadruple $(M,\Pi,R,H)$ consisting of a smooth manifold $M$ equipped with a Poisson structure $\Pi$, an antisymmetric multivector $R$ of degree $p+1$  
	and a closed $(p+2)$-form $H$ such that Eq. \eqref{RPoisson} holds. 
\end{defn}

We note that ``twisted'' in this definition refers to the bracket $[\Pi,R]$, whereas at this stage the bracket $[\Pi,\Pi]$ remains untwisted. Moreover, in case $H=0$ one may refer to the corresponding structure simply as $R$-Poisson structure of order $p+1$. Thus in this subsection we consider that the target space for the class of topological field theories given by the action functional $S^{(p+1)}$ is a twisted $R$-Poisson manifold of order $p+1$. Note that in the way we formulate this definition, a twisted Poisson manifold is \emph{not} a twisted $R$-Poisson manifold of order 2. One may unify these structures by considering a more general definition, however this is not particularly useful for our purposes and therefore we refrain from introducing terminology which is unnecessary for the present paper. Instead we will see how the twisted Poisson structure can play a role in this context in Section \ref{sec4}, where we will present the necessary definition. From the point of view of the present discussion, the structure we consider is closer to ordinary, untwisted Poisson manifolds since it reduces to them when $H=0=R$. However, it still contains a flavour of twisted structures in the sense of \eqref{RPoisson}. 

Returning to the action $S^{(p+1)}$, we are ready to discuss its gauge symmetries. For this purpose, we introduce a $\S_{p+1}$-dependent gauge parameter for each differential form field in the theory, specifically $(\epsilon_i,\chi^{i},\psi_{i})$ of form degrees $(0,p-2,p-1)$ respectively. Consider now the following set of gauge transformations, where for simplicity we have neglected the wedge products between consecutive forms,{\footnote{We note that these transformations are valid also for $p=1$, with the understanding that $Y^{i}$ is a host of scalar fields too, and the parameter $\chi^{i}$ does not exist in that case. Moreover, in this special case one can consider a more general form of these transformations.}} 
\bea 
\label{gt1} \d X^{i}&=&\Pi^{ji}\epsilon_{j}\,,\\[4pt]
\label{gt2} \d A_i&=&\dd\epsilon_i+\partial_i\Pi^{jk}A_j\epsilon_k \,,\\[4pt]
 \d Y^{i}&=&(-1)^{p-1}\dd\chi^{i}+\Pi^{ji}\, \psi_j -\partial_j\Pi^{ik}\left(\chi^{j}A_k+Y^{j}\epsilon_k\right)+\frac 1{(p-1)!}R^{iji_1\dots i_{p-1}}A_{i_1}\dots A_{i_{p-1}}\epsilon_j\,,\nn\\ \\[4pt] \label{gt3}
\d Z_i&=&(-1)^{p}\dd\psi_i+\partial_i\Pi^{jk}\left(Z_j\epsilon_k+\psi_jA_k\right) -\partial_i\partial_j\Pi^{kl}\left(Y^jA_k\epsilon_l-\frac 12 \, A_kA_l\chi^{j}\right) \, + \nn\\ 
&&\qquad \qquad \quad + \, \frac {(-1)^p}{p!} \, \partial_iR^{ji_1\dots i_{p}}A_{i_1}\dots A_{i_{p}}\epsilon_j-\frac 1{(p+1)!}\Pi^{kj}H_{ijl_1\dots l_p}\Omega^{l_1\dots l_p}\epsilon_k \,, \label{gt4}
\eea 
where the $\Omega$ in the last term of the gauge transformation of the highest degree form $Z_i$ is given by the formula  
\bea \label{Omega}
\Omega^{l_1\dots l_p}=\sum_{r=1}^{p+1}(-1)^{r}\prod_{s=1}^{r-1}\dd X^{l_{s}}\prod_{t=r}^{p}\Pi^{l_tm_t}A_{m_t}\,.
\eea 
For the reader's convenience, let us expand this expression to make the structure of the terms in the finite sum more transparent, showing the first two and last two of them:
\bea 
\Omega^{l_1\dots l_p}&=&(-1)^{p+1}\dd X^{l_1}\dots \dd X^{l_p}+(-1)^{p}\Pi^{l_pm_p}\dd X^{l_1}\dots \dd X^{l_{p-1}}A_{m_p}+\dots \, + \nn\\[4pt] && + \, \Pi^{l_2m_2}\dots \Pi^{l_pm_p}\dd X^{l_1}A_{m_2}\dots A_{m_p} -\Pi^{l_1m_1}\dots\Pi^{l_pm_p}A_{m_1}\dots A_{m_p}\,.
\eea 
This rather complicated structure of the last term in the gauge transformation of $Z_i$ is tailored so that the action functional $S^{(p+1)}$ is gauge invariant and it is the higher analogue of the last term in the right hand side of \eqref{HPSMtrafo2} of the HPSM. Indeed, the following proposition holds.

\begin{prop}\label{propSp+1}
	The classical action functional $S^{(p+1)}$ given in \eqref{Sp+1} on a spacetime $\S_{p+1}$ without boundary is invariant under the gauge transformations \eqref{gt1}-\eqref{gt4} if and only if the target space M is a twisted R-Poisson manifold of order p+1 and $\Omega^{l_1\dots l_p}$ is given as in Eq. \eqref{Omega}.  
	\end{prop} 

The proof follows by direct calculations and by taking into account definition \ref{tRPoisson}. For orientation, let us describe some basic aspects of this calculation. 
In the gauge transformed action $\d S^{(p+1)}$ one directly identifies a variety of different potentially ``dangerous'' (i.e. non-vanishing) terms, corresponding to possible ways to form a $(p+1)$-form with the given field content such that each term contains a single gauge parameter $\epsilon_i, \chi^{i}$ or $\psi_i$, and a host of total derivative terms which do not play a role in the classical action since we work on a spacetime without boundary. Schematically, neglecting indices and denoting repeated fields as powers, the combinations $\epsilon Z\dd X, \psi A \dd X, \epsilon YA\dd X, \chi A^2\dd X,\epsilon A^p\dd X,\psi\dd A,\epsilon A\dd Y,\chi A\dd A,\e A^{p-1}\dd A$ and $\epsilon Y \dd A$ cancel directly. In particular one finds 
\bea 
\d S^{(p+1)}&=&\int_{\S_{p+1}}\left\{\epsilon_kZ_i\w A_j\left(3!\Pi^{[k\underline{l}}\partial_l\Pi^{ij]}\right)+\psi_{k}\w A_i\w A_j\left(\frac 12 3! \Pi^{[k\underline{l}}\partial_l\Pi^{ji]}\right)+\right.\nn\\[4pt] 
&&\left.+\, \left(\epsilon_kY^{l}\w A_i\w A_j+\frac 12 \chi^{l}\w A_i\w A_j\w A_k\right)\left(\frac12 \partial_{l}\left(3!\Pi^{[j\underline{m}}\partial_m\Pi^{ik]}\right)\right)+\right.\nn\\[4pt]
&&\left. +\, \epsilon_kA_{k_1}\w\dots\w A_{k_{p+1}}\left(\frac 1{p!}\Pi^{ik_1}\partial_iR^{kk_2\dots k_{p+1}}-\frac 1{(p+1)!}\Pi^{ik}\partial_iR^{k_1\dots k_{p+1}}+\right. \right.
\nn\\[4pt]
&& \left.\qquad \qquad \qquad \qquad \quad \left. +\, \frac 1{p!}R^{ik_2\dots k_{p+1}}\partial_i\Pi^{k_1k}-\frac 1{2(p-1)!}R^{ikk_1\dots k_{p-1}}\partial_i\Pi^{k_pk_{p+1}}\right)+\right.\nn\\[4pt]
&&\left. +\,\frac 1{(p+1)!}\epsilon_k\Pi^{kj}H_{jk_1\dots k_pi}\left(\dd X^{k_1}\w\dots\w\dd X^{k_p}-(-1)^{p+1}\Omega^{k_1\dots k_p}\right)\w\dd X^{i}\, +\right.\nn\\[4pt]
&& \left.+\,\frac{(-1)^{p}}{(p+1)!}\epsilon_k\Pi^{kj}H_{jk_1\dots k_pi}\Omega^{k_1\dots k_p}\w \Pi^{im}A_m \right\}\,,\label{dSintermediate}
\eea 
where antisymmetrizations are taken with weight one and underlined indices do not participate in them. 
We now immediately observe that the first two lines in \eqref{dSintermediate} vanish if and only if $\Pi$ is 
a Poisson 2-vector, which is one of the assumptions in definition \ref{tRPoisson}. Indeed, recall that the Schouten-Nijenhuis bracket in a coordinate basis takes the form 
\be 
[\Pi,\Pi]=\Pi^{il}\partial_l\Pi^{jk}\partial_i\w\partial_j\w\partial_k\,.
\ee
Focusing on the two last lines of \eqref{dSintermediate}, it is straightforward to see that with $\Omega$ given as \eqref{Omega} there is a single term that does not cancel and it is of the form $\epsilon A^{p+1}$, namely it is added in the remaining term appearing in the third and fourth lines. This is the term that contains $p+2$ appearances of $\Pi$ contracted with each of the indices of the $(p+2)$-form $H$. What remains is to recognize that these remaining terms imply the condition \eqref{RPoisson} of Definition \ref{tRPoisson}. To see this, recall the definition of the Schouten-Nijenhuis bracket which renders the space of antisymmetric multivectors a Gerstenhaber algebra,
\be 
[v_1\dots v_p,u_1\dots u_q]=\sum_{i,j}(-1)^{i+j}[v_i,u_j]v_1\dots v_{i-1} v_{i+1}\dots v_{p} u_1\dots  u_{j-1}u_{j+1}\dots u_q\,,
\ee 
where all $v$ and $u$ are vector fields, $[v_i,u_j]$ denotes their ordinary Lie bracket of which the Schouten-Nijenhuis bracket is an extension and the wedge product is implicit in this formula. In a coordinate basis we expand the 2-vector and the $(p+1)$-vector as $\Pi=\frac 12 \Pi^{ij}\partial_i\w\partial_j$ and $R=\frac{1}{(p+1)!}R^{k_1\dots k_{p+1}}\partial_{k_1}\w\dots \w\partial_{k_{p+1}}$ respectively, in which case applying the above formula one finds 
\be 
[\Pi,R]=\frac 1{(p+2)!}\left((p+2)\Pi^{k_1i}\partial_iR^{k_2\dots k_{p}}+\frac{(p+1)(p+2)}{2}R^{k_1ik_2\dots k_{p-2}}\partial_i\Pi^{k_{p-1}k_{p}}\right)\partial_{k_1}\dots\partial_{k_{p+1}}\,.
\ee  
 It is now a simple matter to see that the remaining terms cancel if and only if the condition \eqref{RPoisson} is satisfied, which concludes the proof of the proposition.

\begin{rmk}
	A corollary of Proposition \ref{propSp+1} is that for $p>1$ and in the special case when the Wess-Zumino term and the $(p+1)$-vector $R$ vanish, the target space has the structure of a Poisson manifold. The corresponding topological field theories can therefore be called Poisson sigma models in dimensions $\ge 3$, generalizing the standard Poisson sigma model in 2 dimensions. 
\end{rmk}  

We conclude the analysis in this section with the classical equations of motion for the $(p+1)$-dimensional topological field theory with action functional $S^{(p+1)}$. Variation with respect to each of the four fields yields 
\bea 
\label{eom1} F^{i}&:=&\dd X^{i}+\Pi^{ij}A_j=0\,,
\\[4pt]
\label{eom2} G_{i}&:=&\dd A_i+\frac 12 \partial_i\Pi^{jk}A_j\w A_k=0\,,
\\[4pt]
\label{eom3} {\cal F}^{i}&:=&\dd Y^{i}+(-1)^{p}\Pi^{ij}Z_j+\partial_k\Pi^{ij}A_j\w Y^{k}-\frac 1{p!}R^{ij_1\dots j_p}A_{j_1}\w\dots\w A_{j_{p}}=0\,,
\\[4pt]
{\cal G}_{i}&:=&(-1)^{p+1}\dd Z_i+\partial_i\Pi^{jk}\,Z_j\w A_k-\frac 12 \partial_i\partial_j\Pi^{kl}\,Y^{j}\w A_k\w A_l\, +\nn\\[4pt] 
&& \,+ \,  \frac 1{(p+1)!}\partial_iR^{j_1\dots j_{p+1}}A_{j_1}\w\dots\w A_{j_{p+1}}+\frac 1{(p+1)!}H_{ij_1\dots j_{p+1}}\dd X^{j_1}\w\dots\w \dd X^{j_{p+1}}=0\,. \nn\\ \label{eom4}
\eea 
Evidently, target space covariance is not manifest either in the field equations or in the gauge transformations, much like the HPSM. One difference to the HPSM though is that not even the action itself manifests target space covariance in the form presented so far, since it involves an explicit partial derivative of the Poisson tensor. All these issues are carefully treated in the coming section. 

\section{The target space covariant formulation}
\label{sec3} 

A firm geometric basis for the topological field theories introduced in Section \ref{sec22} requires a target space covariant formulation. This means that we would like to determine the basis independent expressions for the equations of motion, the gauge transformation rules and the action functional itself. This is the purpose of the present section. We note that spacetime covariance is already manifest. 

In order to achieve the above goals, we introduce an auxiliary connection $\mathring\nabla$ on the tangent bundle $TM$. This is an arbitrary connection without torsion, the latter property denoted by a ring over nabla to avoid confusion with the connection $\nabla$ with torsion introduced for the HPSM. As before, the connection coefficients in a holonomic frame are denoted as $\mathring{\Gamma}_{ij}^{k}$, namely $\mathring\nabla\partial_i=\mathring\G_{ij}^{k}\dd x^{j}\otimes \partial_k$, where we recall that $(x^{i})$ are local coordinates on $M$. We also recall that the induced connection of the cotangent bundle $T^{\ast}M$ acts as 
$\mathring\nabla\dd x^{i}=-\mathring{\Gamma}^{i}_{kj}\dd x^{j}\otimes \dd x^{k}$. Since $\mathring\nabla$ is torsionless, the corresponding connection coefficients are symmetric, namely $\mathring\G_{ij}^{k}=\mathring{\Gamma}_{ji}^{k}$. 

Let us now begin examining the covariant form of the field equations given in \eqref{eom1}-\eqref{eom4}. The first one is already covariant as it stands and it may be written in a basis independent form as 
\be 
F=\dd X+\Pi(A)=0\,,
\ee   
where the contraction in the second term is in the second slot as can be seen via \eqref{eom1}. 
In all following formulas, composition with the map $X$ is understood for target space quantities. 
As already mentioned in the two-dimensional case, $\dd X$ is a linear map from $T_{P}\S_{p+1}$ to $T_{X(P)}M$ for every point $P\in \S_{p+1}$. Next, we examine the second classical field equation \eqref{eom2} focusing on determining the target space covariant field strength of the field $A\in \Omega^{1}(\S_{p+1},X^{\ast}T^{\ast}M)$. Using the fact that in local coordinates 
\be 
\mathring{\nabla}_k\Pi^{ij}= \partial_k\Pi^{ij}+\mathring{\G}^{i}_{kl}\Pi^{lj}+\mathring{\G}^{j}_{kl}\Pi^{il}\,,
\ee   
after a little algebra we obtain
\be \label{Ginter}
G_i=\mathring{\DD}A_i+\frac 12 \mathring{\nabla}_i\Pi^{jk}A_j\w A_k-\mathring{\G}^k_{ij}A_k\w F^{j}\,,
\ee  
where $\mathring{\DD}$ is the exterior covariant derivative on forms, induced by the connection $\mathring{\nabla}$. Presently, $\mathring{\DD}A_i=\dd A_i-\mathring{\G}_{ij}^k\dd X^{j}\w A_k$. 
We observe that the last term on the right hand side is proportional to the first field equation and therefore we are prompted to define $G^{\mathring\nabla}_i:=G_i+\mathring{\G}^{k}_{ij}A_k\w F^{j}$. Thus, the corresponding tensor in coordinate independent terms is given as 
\be 
G^{\mathring\nabla}=\mathring{\DD}A-\frac 12\, T(A,A)\,,
\ee 
where we recall that $T=-\mathring{\nabla}\Pi$. 
Then the field equation is equivalent to $G^{\mathring\nabla}=0$. 

The covariantization of the remaining two equations of motion is somewhat more demanding. We continue with the third equation \eqref{eom3}, determining the target space covariant field strength for the field  $Y\in \Omega^{p-1}(\S_{p+1},X^{\ast}TM)$. Using that $\mathring{\DD}Y^{i}=\dd Y^{i}+\mathring{\G}^{i}_{jk}\dd X^{j}\w Y^{k}$, we obtain the intermediate result 
\bea 
{\cal F}^{i}&=&\mathring{\DD}Y^{i}+\mathring{\nabla}_{k}\Pi^{ij}A_j\w Y^{k}+(-1)^{p}\Pi^{ij}(Z_j+\mathring{\G}^{l}_{jk}Y^{k}\w A_l) \, - \nn \\[4pt]&&  \, \qquad  -\, \frac 1 {p!}R^{ij_1\dots j_{p}}A_{j_1}\w\dots \w A_{j_{p}} +(-1)^p \mathring{\G}^{i}_{jk} Y^k\w F^{j}\,.
\eea 
We observe that the last term on the right hand side is once more proportional to $F^{i}$. Moreover, due to the above result, we are prompted to make a field redefinition and introduce the quantity
\be 
{Z}^{\mathring{\nabla}}_i=Z_i+\mathring{\G}_{ij}^{k}Y^{j}\w A_k
\,.\ee 
 In the same spirit as in the previous case, we now define $({\cal F}^{\mathring\nabla})^{i}:={\cal F}^{i}+(-1)^{p-1}\mathring{\G}^{i}_{jk}Y^{k}\w F^{j}$. The final form of the covariant field strength for $Y$ is therefore 
\bea 
{\cal F}^{\mathring\nabla}= \mathring{\DD}Y-T(A,Y)+(-1)^{p}\Pi({Z}^{\mathring{\nabla}})-\frac 1{p!}R(A,\dots,A)\,,
\eea  
where the contractions are inferred from the corresponding local coordinate expression, in particular for the expressions that this might appear ambiguous the rule is $T(A,Y)=T_k^{ij}A_j\w Y^{k} \otimes e^{k}$ and $\Pi(Z^{\mathring{\nabla}})=\Pi^{ij}Z^{\mathring{\nabla}}_j\otimes e_{i}$. 
 
Similar algebraic manipulations apply for the last equation of motion \eqref{eom4}. However, one should pay special attention to its final term involving $H$ and $p+1$ appearances of $\dd X^{i}$. The strategy is to turn each one of them into $F^{i}$, eventually leading to a term proportional to the field equation $F^{i}=0$ and a term with $p+1$ appearances of $A_i$ which should be collected with the penultimate term in ${\cal G}_{i}$. 
Moreover, similarly to the previous cases, the strictly lower-degree field strengths appear on the right hand side of the intermediate result, which in this case means all three field strengths $F, G$ and ${\cal G}$,  suggesting the following definition 
\be 
{\cal G}^{\mathring{\nabla}}_i:={\cal G}_{i}-\mathring{\G}^{k}_{ij}A_k\w{\cal F}^{j}+\mathring{\G}^{k}_{ij}Y^{j}\w G_{k}- M_{il} \w F^{l}\,,
\ee 
where we have defined the following shorthand notation for the $p$-form $M_{il}$ that multiplies the field strength $F$: 
\be 
M_{il}:=\partial_l\mathring{\G}^{j}_{ik}Y^{k}\w A_j- \mathring{\G}_{il}^{k}{Z}^{\mathring{\nabla}}_k-\frac 1{(p+1)!}H_{ilj_1\dots j_{p}}\Omega^{j_1\dots j_p}\,,
\ee 
with $\Omega$ given in \eqref{Omega}. Then we find the following final result for the covariant tensor, 
\bea 
{\cal G}^{\mathring{\nabla}}&=&(-1)^{p+1}\mathring{\DD}{Z}^{\mathring{\nabla}}-T({Z}^{\mathring{\nabla}},A)+\frac 12 \left(\mathring{\nabla}T+2\mbox{Alt}(\iota_{\rho}\mathring{\cal R})\right)(Y,A,A)\, + \nn\\[4pt] 
&& \qquad \qquad \, \quad 
+\, \frac 1{(p+1)!}(\mathring{\nabla}R+{\cal T})(A,\dots,A)\,,
\eea 
where $\text{Alt}$ denotes antisymmetrization over $TM\otimes TM$ and $\rho$ is the vector field of the foliation generated by the Poisson structure $\Pi$. 
Once again, the contraction rules are inferred from the local coordinate result. 
In this expression, $\mathring{\cal R}$ is the Riemann curvature tensor of the connection $\mathring{\nabla}$, reading in components as 
\be 
\mathring{\cal R}^{k}_{lij}=\partial_{i}\mathring{\G}^{k}_{lj}-\partial_j\mathring{\G}^{k}_{li}+\mathring{\G}^{m}_{lj}\mathring{\G}^{k}_{mi}-\mathring{\G}^{m}_{li}\mathring{\G}^{k}_{mj}\,,
\ee 
in terms of the connection coefficients, as usual. Furthermore, the tensor ${\cal T} \in \G(T^{\ast}M\otimes \bigwedge^{p+1}TM)$  is given by
\be 
{\cal T}:=\langle\otimes^{p+1}\Pi,H_{p+2}\rangle\,.
\ee 
The corresponding field equation accompanying the previous ones is then ${\cal G}^{\mathring{\nabla}}=0$. This completes the covariant formulation of all four equations of motion for the topological field theories we consider. 

Equipped with the above results, it is a simple matter to express the action functional \eqref{Sp+1} in a basis independent form, getting rid of the disturbing explicit partial derivative the appears in its fourth term. Certainly there are more than one ways to do that, reflecting essentially the analogy with the HPSM, where we argued that the trivial rewriting as \eqref{SHPSM} or alternatively \eqref{SHPSMalt} at the expense of a sign change in the second term involving the 2-vector has an interesting counterpart for the higher-dimensional topological field theories studied here. Considering this, the fully covariant formulation of the action functional \eqref{Sp+1} may be expressed in the following two ways: 
	\begin{align}
	\label{Sp+1cov}
	 S^{(p+1)} & = \int_{\S_{p+1}}\left(\langle {Z}^{\mathring{\nabla}},F\rangle-\langle Y,G^{\mathring{\nabla}}\rangle+\frac 1{(p+1)!}(R\circ X)(A,\dots,A)\right)+\int_{\S_{p+2}}X^{\ast}H   \\[4pt]
	&= \int_{\S_{p+1}}\left(\langle  Z^{\mathring{\nabla}},\dd X\rangle-\langle A,{\cal F}^{\mathring{\nabla}}\rangle+\frac 12 (\mathring{\nabla}\Pi\circ X)(Y,A,A)\, - \right. \nn\\ 
	&\qquad \qquad \qquad \qquad \qquad \quad \left. - \,   \frac p{(p+1)!} (R\circ X)(A,\dots,A)\right)+\int_{\S_{p+2}}X^{\ast}H\,. \label{Sp+1covb}
	\end{align}
	It is observed that the lower version contains explicitly the covariant derivative of the Poisson tensor $\Pi$ and the field strength of the field $Y$. On the other hand, in the upper version $\Pi$ is completely absorbed in the field strengths of $X$ and $A$ and its covariant derivative does not appear. In any case, the two versions are equal. 

Now that we have found the fully covariant form of the action and the field equations, we turn to the gauge transformations. The one for the scalar fields $X^{i}$ does not require any special treatment since it is given by $\d^{\mathring{\nabla}}(X)=\d X=\Pi(\epsilon)$, where we denoted the covariant transformation rule by $\d^{\mathring{\nabla}}$, which in the present case is identical to the one presented before. This is no longer the case for the rest of the gauge transformations. Starting with \eqref{gt2}, some simple algebra leads to the intermediate result 
\be 
\d A_i-\mathring{\G}^{k}_{ij}\epsilon_k F^{j}+\mathring{\G}^{j}_{il}\Pi^{lk}A_j\epsilon_k=\mathring{\DD}\epsilon_i+\mathring{\nabla}_i\Pi^{jk}A_j\epsilon_k\,.
\ee 
We observe that the right hand side involves only covariant quantities and therefore one should interpret the left hand side appropriately as a covariant transformation rule $\d^{\mathring{\nabla}}A$. Recall now that $A\in \Omega^{1}(\S_{p+1},X^{\ast}T^{\ast}M)$ and choose a basis $e^{i}$ of $X^{\ast}T^{\ast}M$, in which case one may write $A=A_i\otimes e^{i}$. The frame itself changes according to $\d^{\mathring{\nabla}}e^{i}=-\mathring{\G}^{i}_{jk}\d X^{j}e^{k}$ and applying the Leibniz rule one finds  
\be 
\d^{\mathring{\nabla}}A=\d^{\mathring{\nabla}}A_i\otimes e^{i}+A_{i}\otimes \d^{\mathring\nabla}e^{i}=(\d^{\mathring{\nabla}}A_i+\mathring{\G}^{j}_{il}\Pi^{lk}A_j\epsilon_k)\otimes e^{i}\,. 
\ee 
This indicates that the covariant gauge transformation for the components $A_i$ should be promoted to 
$\d^{\mathring{\nabla}}A_i=\d A_i-\mathring{\G}^{k}_{ij} \epsilon_kF^{j}$, in which case one concludes that 
\be 
\d^{\mathring{\nabla}}A=\mathring{\DD}\epsilon-T(A,\epsilon)\,,
\ee 
this being the desired covariant result. An equivalent way of obtaining this already appears in \cite{Ikeda:2019czt} in the context of the HPSM.

Next, we should find the covariant counterpart for the gauge transformation \eqref{gt3} of the field $Y\in \Omega^{p-1}(\S_{p+1},X^{\ast}TM)$. We follow the same strategy as in the previous case, namely we first promote the exterior derivative $\dd$ on the gauge parameter to the covariant one $\mathring{\DD}$ and express the partial derivative acting on the Poisson tensor in terms of the covariant one with respect to the connection $\mathring{\nabla}$. This leads to the intermediate result 
\bea 
\d Y^{i}-\mathring{\G}^{i}_{jk}\chi^{j}\w F^{k}-\mathring{\G}^{i}_{jk}\Pi^{kl}Y^{j}\epsilon_l&=&(-1)^{p-1}\mathring{\DD}\chi^{i}+\Pi^{ji}{\psi}^{\mathring{\nabla}}_{j}-\mathring{\nabla}_j\Pi^{ik}(\chi^{j}\w A_k+Y^{j}\epsilon_k)+\, \nn\\[4pt] && +\,  \frac 1{(p-1)!}R^{ijk_1\dots k_{p-1}}A_{k_1}\w\dots \w A_{k_{p-1}}\epsilon_j\,,
\eea  
where we have redefined the $(p-1)$-form gauge parameter $\psi_i$ to ${\psi}^{\mathring{\nabla}}_i:=\psi_i-\mathring{\G}_{ij}^{k}(\chi^{j}\w A_k+Y^{j}\epsilon_k)$, in accord with the field redefinition from the field $Z_i$ to the new field ${Z}^{\mathring{\nabla}}_{i}$. Once more we observe that the right hand side is a covariant expression and therefore it should be interpreted as the covariant gauge transformation $\d^{\mathring{\nabla}}Y$. This turns out to be the case when we introduce a dual basis $e_i$ of $X^{\ast}TM$ and write $Y=Y^{i}\otimes e_{i}$. The dual basis transforms as $\d^{\mathring{\nabla}}e_i=\mathring{\G}_{ij}^{k}\d X^{j}e_{k}$ and therefore using the Leibniz rule one finds 
\be 
\d^{\mathring{\nabla}}Y= \d^{\mathring{\nabla}}Y^{i}\otimes e_i+Y^{i}\otimes \d^{\mathring{\nabla}}e_i= (\d^{\mathring{\nabla}}Y^{i}+\mathring{\G}^{i}_{jk}\Pi^{lk}Y^{j}\epsilon_l)\otimes e_i\,.
\ee 
This means that the gauge transformation $\d Y^{i}$ for the components $Y^{i}$ should be promoted to $\d^{\mathring{\nabla}}Y^{i}=\d Y^{i}-\mathring{\G}^{i}_{jk}\chi^{j}\w F^{k}$, in which case the final form of the covariant gauge transformation for the full field $Y$ is given as 
\bea 
\d^{\mathring{\nabla}}Y&=&(-1)^{p-1}\mathring{\DD}\chi +\Pi({\psi}^{\mathring{\nabla}})+T[(\chi,A)+(Y,\epsilon)] + \frac 1{(p-1)!}R(\epsilon,A,\dots,A)\,.  
\eea 
Finally, we should determine the covariant form of the fourth gauge transformation \eqref{gt4}. From the previous discussion, it should be clear that we are looking for the transformation of the redefined field ${{Z}^{\mathring{\nabla}}}\in \Omega^{p}(\S_{p+1},X^{\ast}T^{\ast}M)$ in terms of the redefined $(p-1)$-form gauge parameter ${\psi}^{\mathring{\nabla}}$, since this is the one that appears in all target space covariant formulas until now. The strategy to find $\d^{\mathring{\nabla}}{Z}^{\mathring{\nabla}}$ follows the same steps as previously, albeit with significantly more complicated algebraic manipulations due. After a calculation, one ends up with the following intermediate formula
\bea 
&& \d{Z}^{\mathring{\nabla}}_{i}+\mathring{\G}^{j}_{il}\Pi^{lk}{Z}^{\mathring{\nabla}}_{j}\epsilon_{k}-\mathring{\G}^{k}_{ij}\chi^{j}\w G_k-(-1)^{p}\mathring{\G}^{k}_{ij}{\cal F}^{j}\epsilon_k-(-1)^{p}F^{j}\w N_{ij}= \nn\\[4pt] 
&& = (-1)^{p}\mathring{\DD}{\psi}^{\mathring{\nabla}}_i +\mathring{\nabla}_{i}\Pi^{jk}{\psi}^{\mathring{\nabla}}_j\w A_k+\mathring{\nabla}_i\Pi^{jk}{Z}^{\mathring{\nabla}}_{j}\epsilon_k\, + \nn\\[4pt] && +\,  \frac{(-1)^{p}}{p!}\left(\mathring{\nabla}_iR^{jl_1\dots l_p}+{\cal T}_i^{jl_1\dots l_p}\right)A_{l_1}\w\dots \w A_{l_{p}}\epsilon_j \, - \nn\\[4pt] && - \left(\mathring{\nabla}_i\mathring{\nabla}_l\Pi^{mk}+2\Pi^{j[m}\mathring{\cal R}^{k]}_{lij}\right)\left(Y^{l}\w A_m\epsilon_k-\frac 12 \chi^{l}\w A_m\w A_k\right)\,,
\eea  
where we observe that the left hand side contains all lower-degree field strengths and moreover we have defined a shorthand notation for the spacetime $(p-1)$-form that multiples the field strength $F$:
\be 
N_{ij}:=\mathring{\G}^{k}_{ij}{\psi}^{\mathring{\nabla}}_k+\left(\partial_j\mathring{\G}_{il}^{m}+\mathring{\G}_{ij}^{k}\mathring{\G}_{kl}^{m}\right)(\chi^{l}\w A_m+Y^{l}\epsilon_m)+\frac 1{(p+1)!}\Pi^{mm_p}\epsilon_mH_{ijm_1\dots m_p}\widetilde{\Omega}^{m_1\dots m_{p-1}}\,,
\ee 
with $\widetilde{\Omega}$ being a7 $(p-1)$-form given by the formula (note that this is different from $\Omega$ defined in a similar way through \eqref{Omega}) 
\be \label{Omegatilde}
\widetilde{\Omega}^{m_1\dots m_{p-1}}:=\sum_{r=0}^{p-1}(-1)^{r}(p-r)\prod_{s=1}^{r}\dd X^{m_s}\prod_{t=r+1}^{p-1}\Pi^{m_tl_t}A_{l_t}\,.
\ee 
It is immediately observed that the above manipulation has led to a covariant expression on the right hand side, which should be identified as $\d^{\mathring{\nabla}}{Z}^{\mathring{\nabla}}$. Since in the basis $e^{i}$ of the pull-back bundle $X^{\ast}T^{\ast}M$, we find that  $\d^{\mathring{\nabla}}{Z}^{\mathring{\nabla}}=(\d^{\mathring{\nabla}}{Z}^{\mathring{\nabla}}_i-\mathring{\G}_{ij}^{k}\d X^{j}\mathring{Z}_{k})\otimes e^{i}$. This leads to the improved covariant transformation rule 
\be 
\d^{\mathring\nabla} {Z}^{\mathring{\nabla}}_{i}=\d {Z}^{\mathring{\nabla}}_{i}-\mathring{\G}^{k}_{ij}\chi^{j}\w G_k-(-1)^{p}\mathring{\G}^{k}_{ij}{\cal F}^{j}\epsilon_k-(-1)^{p}F^{j}\w N_{ij}\,.
\ee 
Finally, the resulting covariant gauge transformation for the field ${Z}^{\mathring{\nabla}}$ is 
\bea 
\d^{\mathring{\nabla}}{Z}^{\mathring{\nabla}}&=&(-1)^{p}\mathring{\DD}{\psi}^{\mathring{\nabla}}-T[({\psi}^{\mathring{\nabla}},A)+({Z}^{\mathring{\nabla}},\epsilon)]+\frac{(-1)^{p}}{p!}(\mathring{\nabla}R+{\cal T})(\epsilon,A,\dots,A) \nn\\[4pt] &&\qquad \qquad \, + \, \left(\mathring{\nabla}T+2\mbox{Alt}(\iota_{\rho}\mathring{\cal R})\right)[(Y,A,\epsilon)-\frac 12 (\chi,A,A)]
\,.
\eea 

Summarizing the discussion on the target space covariant form of the gauge transformation rules of the fields and of the corresponding field strengths that define the classical equations of motion for the topological field theories given by the action functional $S^{(p+1)}$, we collect our results in Table \ref{table1}.

\begin{table}
\begin{tabular}{ |c||c|c| }
	\hline
	\multirow{5}{3.5em}{Field $\Phi$} && \\ & &  Gauge transformation $\d^{\mathring{\nabla}}\Phi$ \\  & $\in$ & \\ & & Field strength ${\mathbb F}^{\mathring{\nabla}}$ \\ &&  \\ 
	\hline  
	\multirow{5}{1em}{$X$} &&\\  & &  $\Pi(\epsilon)$ \\ & ${\cal C}^{\infty}(\S)$    &     \\ 
	& &  $\dd X+\Pi(A)$ \\ && \\ 
	\hline 
	\multirow{5}{1em}{$A$}&&\\  && $\mathring{\DD}\epsilon-T(A,\epsilon)$\\  &   $\Omega^{1}(\S,X^{\ast}T^{\ast}M)$  &  \\ & & $\mathring{\DD}A-\frac 12 \, T(A,A)$ \\ && \\ \hline 
	\multirow{5}{1em}{$Y$}&&\\ & & $(-1)^{p-1}\mathring{\DD}\chi +\Pi({\psi}^{\mathring{\nabla}})+T[(\chi,A)+(Y,\epsilon)]+\frac 1{(p-1)!}R(\epsilon,A,\dots,A)$  \\ & $\Omega^{p-1}(\S,X^{\ast}TM)$ &  \\ & & $\mathring{\DD}Y-T(A,Y)+(-1)^{p}\Pi({Z}^{\mathring{\nabla}})-\frac 1{p!}R(A,\dots,A)$ \\ &&\\ \hline 
	  \multirow{7}{1em}{${Z}^{\mathring{\nabla}}$} && \\ &&  $(-1)^{p}\mathring{\DD}{\psi}^{\mathring{\nabla}}-T[({\psi}^{\mathring{\nabla}},A)+({Z}^{\mathring{\nabla}},\epsilon)]+\frac{(-1)^{p}}{p!}(\mathring{\nabla}R+{\cal T})(\epsilon,A,\dots,A)$ \\ &&  $+  \left(\mathring{\nabla}T+2\mbox{Alt}(\iota_{\rho}\mathring{\cal R})\right)[(Y,A,\epsilon)-\frac 12 (\chi,A,A)]$
	   \\  &&\\ & $\Omega^{p}(\S,X^{\ast}T^{\ast}M)$ & \\ & & $(-1)^{p+1}\mathring{\DD}{Z}^{\mathring{\nabla}}-T({Z}^{\mathring{\nabla}},A)+\frac 12 \left(\mathring{\nabla}T+2\mbox{Alt}(\iota_{\rho}\mathring{\cal R})\right)(Y,A,A)+ $ \\ && \vspace{-8pt}\\ & & $+\frac 1{(p+1)!}(\mathring{\nabla}R+{\cal T})(A,\dots,A)$ \\  & & \\
	\hline
\end{tabular} \caption{Covariant gauge transformations and field strengths ${\mathbb F}^{\mathring{\nabla}}=(F^{\mathring{\nabla}},G^{\mathring{\nabla}},{\cal F}^{\mathring{\nabla}},{\cal G}^{\mathring{\nabla}})$ for the collection of fields $\Phi=(X,A,Y,{Z}^{\mathring{\nabla}})$ on $\S\equiv \S_{p+1}$. We have defined $T=-\mathring{\nabla}\Pi$. Composition of all geometric background data with the basic map $X:\S_{p+1}\to M$ is understood in all formulas, for instance $\Pi(A)$ is $(\Pi\circ X)(A)$ etc.}\label{table1}
\end{table} 

\section{The Q-manifold picture}
\label{sec4}

Having established the existence of topological field theories with twisted $R$-Poisson structure and their fully covariant formulation, we would now like to understand the structure of the target space as a graded supermanifold, i.e. as a supermanifold endowed with an additional $\Z$-grading in its algebra of functions. A $Q$-manifold---or differential graded (dg) manifold---is a graded supermanifold equipped with a (co)homological vector field $Q$ of degree 1, (co)homological meaning that it squares to zero, namely $Q^{2}=0$. It is well-known that algebroid structures find a realization in terms of $Q$-manifolds. In the simplest case, a Lie algebroid $(E,[\cdot,\cdot]_{E},\rho:E\to TM)$ over $M$, consisting of a vector bundle $E\overset{\pi}\to M$, a Lie algebra bracket on its sections and a smooth (anchor) map from $E$ to the tangent bundle which generates a homomorphism of bundles, may be completely characterised as follows. Consider the parity-reversed vector bundle $E[1]$, which may be described by means of local coordinates $x^{i}$ and $\xi^{a}$ of degree zero (``bosonic'') and one (``fermionic'') respectively, the index $a$ being with respect to a basis $\mathfrak{e}^{a}$ of the dual vector bundle $E^{\ast}$. Equip $E[1]$ with the following degree-1 vector field 
\be 
Q_{E}=\rho^{i}_{a}(x)\xi^{a}\partial_{x^{i}}-\frac 12 C^{a}_{bc}(x)\xi^{b}\xi^{c}\partial_{\xi^{a}}\,,
\ee 
where $\partial_{x^{i}}:=\partial/\partial x^{i}$ and $\partial_{\xi^{a}}:=\partial/\partial{\xi^{a}}$. Then, as first shown by Vaintrob \cite{Vaintrob},  $Q_E^{2}=0$ results in the defining properties of a Lie algebroid provided we make the identifications $\rho(\mf{e}_a)=\rho_{a}^{i}(x)\partial_{x^{i}}$ for the anchor and $[\mf{e}_a,\mf{e}_b]_{E}=C_{ab}^{c}(x)\mf{e}_c$ for the Lie bracket in a local basis $\mf{e}_a$ of $E$. We note that in this graded picture, signs follow the degree of the respective coordinates, specifically for any two coordinates of definite degree
\be 
\xi \eta= (-1)^{|\xi|\eta|}\eta \xi\,.
\ee 
This is also used below, where more degrees appear, other than just 0 and 1.  

Recognizing that a twisted Poisson structure $(M,\Pi,H)$ induces a Lie algebroid structure on the cotangent bundle $T^{\ast}M$, one may use the above construction to show that the associated $Q$-manifold is $(T^{\ast}[1]M,Q_{T^{\ast}M})$. The graded cotangent bundle is equipped with coordinates $x^{i}$ and $\xi_i$ and the desired homological vector field is 
\be \label{Qhp}
Q_{T^{\ast}M}=\Pi^{ij}(x)\xi_i\partial_{x^{j}}-\frac 12 (\partial_i\Pi^{jk}+\Pi^{jl}\Pi^{km}H_{ilm})\xi_{j}\xi_{k}\partial_{\xi_{i}}\,.
\ee     
The correspondence is that $Q_{T^{\ast}M}$ is homological if and only if the 2-vector $\Pi$ satisfies the defining condition \eqref{twistedPoisson} of a twisted Poisson structure \cite{Ikeda:2019czt}. We note in passing that a more covariant version of \eqref{Qhp} is 
\be \label{Qhpb}
Q_{T^{\ast}M}=\Pi^{ij}\xi_i(\partial_{x^{j}}+\Gamma^{l}_{kj}\xi_l\partial_{\xi_k})-\frac 12 \mathring{\nabla}_i\Pi^{jk}\xi_j\xi_k\partial_{\xi_i}\,,
\ee 
where the components of the 3-form $H$ appear only in the supercovariant derivative of the first term through the coefficients of the connection $\nabla$.

A natural question then is \emph{what is the $Q$-manifold associated to the twisted $R$-Poisson structure underlying the topological field theories discussed before?} To answer this question, we first should determine the corresponding graded manifold.{\footnote{In the rest of this section, the word manifold is taken to mean supermanifold.}} The strategy is to follow the field content of the theory. Indeed, the HPSM contained the fields $X^{i}$ and $A_{i}$, which may be seen as pull-backs of the coordinates $x^{i}$ and $\xi_{i}$ of $T^{\ast}[1]M$ by means of two maps $X$ and $A$. The former map is simply the base sigma model map $X:\S_2\to M$. On the other hand, one may now consider the spacetime itself as a graded manifold $T[1]\Sigma_2$, in which case the second map is{\footnote{As in the case of $X$, we take the liberty of denoting this map with the same symbol as the field $A\in\Omega^{1}(\S_{2},X^{\ast}T^{\ast}M)$, as long as no confusion arises.}} $A:T[1]\S_{2}\to T^{\ast}[1]M$. Then we recognize that 
\be 
X^{i}=X^{\ast}(x^{i}) \quad \text{and} \quad A_i=A^{\ast}(\xi_i)\,.  
\ee 
In other words, as long as we know the field content of our theory, we can associate it to a graded manifold accordingly. Let us now return to the general case, namely the one with field content $(X^{i},A_i,Y^{i},Z_i)$ of form degrees $(0,1,p-1,p)$ respectively. The above discussion suggests that we should introduce four sets of coordinates with corresponding degrees on the graded target space, which we denote as $(x^{i},a_i,y^{i},z_{i})$ to be aligned with the notation of the fields in the theory. In which graded manifold can these coordinates arise in a local patch? The rather obvious answer is that the graded target space should be taken to be 
\be 
{\cal M}=T^{\ast}[p]T^{\ast}[1]M\,.
\ee 
Note that utilizing the isomorphism of the tangent and cotangent bundles on $M$ one may relate ${\cal M}$ to the graded manifold $T^{\ast}[p]T[1]M$, which is more often encountered in the literature. However, although for $p=2$ the field content remains essentially the same, for $p>2$ one should note that while for ${\cal M}$ the ``middle'' fields $A$ and $Y$, which are of different degree, take values in $T^{\ast}M$ and $TM$ respectively, for $T^{\ast}[p]T[1]M$ the bundles where they take values are exchanged. For this reason, the type of topological field theories we present here truly require the above choice of ${\cal M}$ in order to be realised. This was also discussed in a special four-dimensional case in \cite{Chatzistavrakidis:2019seu}. The fields of the theory are now obtained as pull-backs of the four types of graded coordinates according to 
\be 
X^{i}=X^{\ast}(x^{i})\,, \quad A_{i}=A^{\ast}(a_i)\,, \quad Y^{i}=Y^{\ast}(y^{i}) \quad \text{and} \quad Z_{i}=Z^{\ast}(z_i)\,,
\ee   
where the maps $X, A, Y$ and $Z$ (once more, not to be confused with the corresponding fields)
are the components of a ``big map'' 
\be 
\Phi: T[1]\S_{p+1} \to {\cal M}\,.
\ee  
This is the augmented sigma model map which may be used when one addresses the AKSZ construction of the models considered here. For the purposes of the present section, we proceed with the discussion of the $Q$-structure on the graded manifold ${\cal M}$. 
Based on the above discussion, we propose the following:
\begin{prop}\label{Qp}
	The pair $(T^{\ast}[p]T^{\ast}[1]M,Q)$ with the degree-1 vector field given by 
	\bea 
	Q&=&\Pi^{ji}a_j\partial_{x^{i}}-\frac 12 \partial_i\Pi^{jk}a_ja_k\partial_{a_{i}}+\left((-1)^{p}\Pi^{ji}z_j-\partial_j\Pi^{ik}a_ky^{j}+\frac 1{p!}R^{ij_1\dots j_p}a_{j_1}\dots a_{j_p}\right)\partial_{y^{i}} \, + \nn\\[4pt] 
	&& +\, \left(\partial_i\Pi^{jk}a_kz_j-\frac {(-1)^{p}}2 \partial_i\partial_j\Pi^{kl}y^{j}a_ka_l+\frac {(-1)^{p}}{(p+1)!} f_{i}^{k_1\dots k_{p+1}}a_{k_1}\dots a_{k_{p+1}}\right)\partial_{z_{i}}\,,\label{Qp+1}
	\eea 
	where $f_{i}^{k_1\dots k_{p+1}}= \partial_iR^{k_1\dots k_{p+1}}+\prod_{r=1}^{p+1}\Pi^{k_rl_r}H_{il_1\dots l_{p+1}}$
	is a $Q$-manifold if and only if $(M,\Pi,R,H)$ is a twisted $R$-Poisson manifold of order $p+1$.  
\end{prop}

To prove the above assertion, it is enough to show that the conditions stemming from $Q^{2}=0$ are the same as the conditions for a twisted $R$-Poisson structure of order $p+1$, namely that $\Pi$ is a Poisson 2-vector and along with the $(p+1)$-vector $R$ and the $(p+2)$-form $H$ they satisfy the fundamental property \eqref{RPoisson}. 
This can be shown by direct calculation of the involved quantities and indeed one easily finds 
\bea 
Q^{2}x^{i}=0 \quad \Leftrightarrow \quad [\Pi,\Pi]=0\,.
\eea 
Moreover, $Q^{2}A_i=\frac 12 \partial_i(\Pi^{lj}\partial_l\Pi^{km})a_ja_ka_m$, which is an identity as long as $\Pi$ is Poisson. Crucially, having established that $\Pi$ is Poisson, we find by direct calculation that  
\be 
Q^{2}Y^{i}=0 \quad \Leftrightarrow \quad  [\Pi,R]=(-1)^{p+2}\langle \otimes^{p+2}\Pi,H_{p+2}\rangle
\ee  
The remaining condition $Q^{2}Z_i$ is then identically satisfied. 

We close this section with a brief discussion on the symplectic structure of the above $Q$-manifolds. Recall that a differential graded symplectic manifold of degree $n$, or $QPn$ manifold in short, is a $Q$ manifold with a compatible symplectic ($P$) structure of degree $n$. Compatibility is expressed through the condition 
\be \label{compatibility}
{\cal L}_{Q}\omega=0\,,
\ee  
where $\omega$ is the symplectic form that defines the $P$ structure on the graded supermanifold. $QP$ manifolds are instrumental in the AKSZ construction of topological field theories \cite{Alexandrov:1995kv}. On the other hand, twists such as the ones discussed in the present paper obstruct $QP$-ness, specifically the compatibility between the $Q$ and $P$ structures. Therefore, the AKSZ construction is not directly applicable to such twisted topological field theories, which explains why the BV-BRST action for the twisted Poisson sigma model in two dimensions was only found recently \cite{Ikeda:2019czt}, although the one of the ordinary PSM can be found in a straightforward way using AKSZ. This is also the reason that we will not introduce a Hamiltonian function $\Theta$, which in the compatible case controls the odd vector field $Q$ via 
$Q=\{\Theta,\cdot\}$, the bracket being the graded Poisson one defined through the graded symplectic structure $\omega$. In that case, the odd vector field is homological if and only if the Maurer-Cartan equation $\{\Theta,\Theta\}=0$ holds. 

In this paper, we are not going to discuss the BV action of the higher-dimensional twisted field theories described here, since this requires a special treatment due to the much higher complexity than the two-dimensional case. In any case, since the $QP$ structure is obstructed as a general rule for models with Wess-Zumino term, $QP$ does not play an equally fundamental role in that case. Nevertheless, it is worth discussing the $QP$ structure of the untwisted models, since this can be used to apply the AKSZ construction to them. First, for the ordinary Poisson structure with homological vector field given in \eqref{Qhp} with $H=0$, the canonical symplectic form is of degree 1 and reads in Darboux coordinates as 
\be 
\omega=\mathtt{d} x^{i}\w \mathtt{d}\xi_{i}\,,
\ee     
where we use the typewriter font $\mathtt{d}$ for the differential on $M$, not to be confused with the one on $\S$ denoted as $\dd$. Since $\mathtt{d}\omega=0$, it is simple to see that \eqref{compatibility} is satisfied (for $H=0$.) In addition, a Hamiltonian is the degree-2 homological function $\Theta=\frac 12 \Pi^{ij}\xi_i\xi_j$ and indeed one may easily find that $\{\Theta,\cdot\}$ is equal to the odd vector field \eqref{Qhp}, or equivalently \eqref{Qhpb}. 

Similarly, consider the $Q$ manifold of Proposition \eqref{Qp} with homological vector field given by \eqref{Qp+1} and assume further that $H_{p+2}=0$. This brings us to the case of an untwisted $R$-Poisson structure. The canonical symplectic form is now of degree $p$ and it reads as 
\be \label{Pp+1}
\omega=\mathtt{d}x^{i}\w \mathtt{d}z_i+\mathtt{d}a_{i}\w\mathtt{d}y^{i}\,,
\ee  
in terms of the coordinates introduced above. Once more, $\mathtt{d}\omega=0$ and a straightforward calculation establishes that the compatibility condition \eqref{compatibility} is satisfied for $H_{p+2}=0$. Therefore we conclude that the triple $(T^{\ast}[p]T^{\ast}[1]M,\omega,Q)$ with $P$ and $Q$ structures given by \eqref{Pp+1} and \eqref{Qp+1} respectively is a $QP$ manifold of degree $p$ if and only if $(M,\Pi,R)$ is a (untwisted) $R$-Poisson manifold of order $p+1$. 

In the general case, where $H_{p+2}\ne 0$, the compatibility condition is not satisfied. One could address this issue in the context of twisted $QP$ manifolds, introduced in \cite{Ikeda:2013wh}. Furthermore, twisted Poisson structures may be also understood in the context of $L_{\infty}$ algebras and homotopy Poisson structures, as in \cite{Roytenberg:2007zz} and \cite{Lang}.  It would be interesting to adopt this perspective also for the twisted $R$-Poisson structures studied here. As already mentioned, such $QP$ structures are mainly useful in finding the BV action of the associated topological field theories and the situation becomes more complicated when a vanilla $QP$ structure is not available. 
A more direct route to determine the BV action is then necessary, which will be applied in \cite{prep}, and therefore we are not going to discuss twisted $QP$ structures further.

\section{Deformations, bi-twisted R-Poisson structures and examples}
\label{sec5}

In previous sections, we assumed from the beginning that $\Pi$ is a Poisson 2-vector and therefore that the target space is equipped with a twisted $R$-Poisson structure. We would now like to ask whether there exist topological field theories other than the two-dimensional HPSM where the target space is equipped with a twisted Poisson structure instead, in presence of the multivector $R$ and the Wess-Zumino term $H$.
One might think that since $\Pi$ ceases to be a Poisson 2-vector its Schouten-Nijenhuis bracket with itself should be  controlled by the available $(p+2)$-form $H$. Although this is what happens in the HPSM, it is not a general feature and it is not what we study here.{\footnote{In fact, it appears impossible to construct such field theories other than the HPSM with the given field content that we assume in this paper, although we will not provide a proof of this negative statement here.}} Instead we determine possible additional terms that can be included in the action functional $S^{(p+1)}$ with the given field content in special dimensions. This will reveal how the HPSM itself is obtained in our approach but also that a three-dimensional topological field theory with twisted Poisson structure exists too, with further accompanying structures that we identify.

\subsection{Special cases and deformations}
\label{sec51}

The approach we followed in defining the topological field theories given by $S^{(p+1)}$ was very general, nevertheless one can identify two shortcomings. The first is that as it stands, the HPSM is \emph{not} included in these theories. The second is that there may exist special cases where with the given field content one can construct more general topological field theories than the ones we presented. In this section we address both above issues and show how they can be rectified. 

Recall that the field content of the theories comprises differential forms $(X,A,Y,Z)$ on $\S_{p+1}$ of degrees $(0,1,p-1,p)$. Focusing on the last three, we have included in $S^{(p+1)}$ three terms, which are of the form $ZA$, $YA^{2}$ and $A^{p+1}$, where the superscripts in this notation denote powers. These are the only three terms of degree $p+1$ that exist for every $p$. However, for specific values of $p$ this may not be the case. Let us determine which are these specific values. 
First, one may ask which terms involving only one of the fields $A, Y$ or $Z$ are admissible. Starting with the $p$-form $Z$, suppose we wish to write a term of type $Z^{q}$. Since this should be a $(p+1)$-form on spacetime, $pq=p+1$ should hold. This means that $p=1/(q-1)$ and of course $p\in \Z$. The only possibility is evidently $q=2$, in which case $p=1$. Therefore, one may include a $Z^{2}$ term in the two-dimensional theory only. As we will see below, this accounts precisely for the HPSM. Similarly, in order to have a $Y^{q}$ term, the condition is $p=(q-1)/(q+1) \in\Z$ for $q>1$. The only solutions are $p=3$ for $q=2$ and $p=2$ for $q=3$, since any other $q$ does not result in an integer $p$. Since the $A^{q}$ term is already accounted for in the general theory, this completes this part of the analysis. 

Next, we ask whether further terms involving exactly two fields are admissible. Studying the possibility of type $Z^qY^r$ it quickly becomes clear that there are two options: a term $ZY$ in three dimensions ($p=2$) and a term $Z^{2}Y$ in two dimensions ($p=1$). In the latter case, $Y$ is a scalar and essentially this is already considered above, therefore we discard it. Moreover, terms of type $Z^qA^r$ are not admissible, save for the generic one considered in previous sections. The last option is $Y^qA^r$, which yields two possibilities, namely a term $Y^2 A$ in three dimensions and a term $Y^qA^2$ in two dimensions. Once more, the latter can be neglected, since it is already considered as long as the corresponding $R^{ij}$ is a function of both $X$ and $Y$ which are the scalar fields in the two dimensional version of the theory. Finally, one can easily find that there is no admissible $Z^qY^rA^s$ term for non-vanishing powers. All terms uncovered here may be thought of as deformations of the general model given by $S^{(p+1)}$, giving rise to special topological field theories that extend the general one in specific dimensions. This is discussed in the ensuing. The above discussion is summarized in Table \ref{table2} for readability. 
\begin{table}\centering
	\begin{tabular}{ |c||c| }
		\hline
		{$\text{dim}\, \S_{p+1}$} & Admissible deformations  \\ 
		\hline  
		\multirow{3}{0.5em}{2} &\\ & $f^{ij}(X,Y)Z_i\w Z_j$, \, $f^{ij}(Y)A_i\w A_j$  \\  & \\
		\hline 
			\multirow{3}{0.5em}{3} &\\ & $f_{ijk}(X)Y^{i}\w Y^{j}\w Y^{k}$, \, $f^{i}_{j}(X)Z_i\w Y^{j}$,\, $f_{ij}^{k}(X)Y^{i}\w Y^{j}\w A_k$\\ &\\ \hline 
	 	\multirow{3}{0.5em}{4} &\\ &  $f_{ij}(X)Y^i\w Y^j$ \\& \\ \hline 
	\end{tabular} \caption{Admissible deformations of the general action functional $S^{(p+1)}$ for special values of $p$. All deformations appear with a suitable function of the scalar fields of the theory, which we denote collectively by $f$ (each being a different function). We observe that special cases appear only in low dimensions $\le 4$.}\label{table2}
\end{table}

 \subsection{Examples and islands of (bi-)twisted R-Poisson TFTs}
 \label{sec52}
 
 We are now in position to discuss examples of topological field theories of the kind introduced in Section \ref{sec2} and also beyond these, in the sense of Section \ref{sec51}. We focus on two, three and four dimensions. Rather than simply reiterating the general case given by $S^{(p+1)}$, which exists in any dimension of $\S_{p+1}$, we highlight the more general theories that include the admissible deformations classified in Table \ref{table2}, which only exist in low dimensions. This will allow us to define not only theories with twisted $R$-Poisson structure but also ``islands'' of  theories such as ones with bi-twisted $R$-Poisson one, as defined below.  
 
 Much like in Section \ref{sec2}, we begin our analysis in local coordinates for simplicity; the upgrade to a coordinate and bases-independent formulation will quickly follow. The set up is essentially the same as before, in that we formally consider again the action \eqref{Sp+1}, albeit with the difference that the function $Q_k^{ij}$ is not taken to be  the partial derivative of the 2-vector components any longer.   
 
 \subsubsection{Twisted R-Poisson 2D TFT and the HPSM}
 \label{sec522}
 
 First we discuss the two-dimensional examples, including the admissible deformations. 
 For clarity, we work in terms of local coordinates in the target space; the corresponding covariant expressions can be found by the method demonstrated in Section \ref{sec22}. The full action functional including the deformations with suitable background functions becomes 
 \bea 
 S^{(2)}_+ &=& S^{(2)}+S^{(2)}_{\text{def}} \nn\\[4pt] 
 &=& \int_{\S_{2}}\left(Z_i\w\dd X^{i}-A_i\w \dd Y^{i}+f_1^{ij}(X,Y)\, Z_i\w A_j+ \frac 12 f_2^{ij}(X,Y)A_i\w A_j \,+   \right. \nn \\[4pt] && \quad \qquad \left. +\,\frac 12  f_3^{ij}(X,Y) Z_i\w Z_j\right)+\int_{\S_{3}}H_3(X,Y)\,. \label{Sp+1+}
 \eea 
 Note that for this case, since $p=1$, the field content comprises \emph{two} sets of scalar fields $X^{i}$ and $Y^{i}$ and two spacetime 1-forms $A_i$ and $Z_i$. We have thus allowed dependence of all background data on both scalar fields, including the 3-form $H_3$. We have modified the notation with respect to previous sections in order to be able to discuss different limits in a proper way, thus we introduced the background data $f_{s}, s=1,2,3$. 
 
 First let us see how this model can be related to the HPSM of Section \ref{sec21}.  This is achieved by 
 choosing $f_1=0=f_2$ and $f_3=X^{\ast}\Pi$, $\Pi$ being an $H$-twisted Poisson 2-vector, and with Wess-Zumino term depending only on $X$, namely being $X^{\ast}H_3$. Then one can immediately see that 
 \be 
 S^{(2)}_+[X,Y,A,Z]|_{f_1=f_2=0, f_3=\Pi}=S_{\text{HPSM}}[X,Z]+S_{\text{BF}}[Y,A]\,,
 \ee  
 namely we obtain the HPSM trivially coupled to an Abelian BF theory in two dimensions. The gauge transformations are the usual ones of the corresponding theories. 
 This is then how the HPSM is included as a special case of the class of topological field theories we consider. 
 
 Clearly, $S^{(2)}_+$ is more general and it includes a twisted $R$-Poisson structure of order 2, which as we have already discussed is distinct from a twisted Poisson structure. This is obtained by the choice 
 $f_3=0$, $f_1(X,Y)=X^{\ast}\Pi$ with $\Pi$ being a Poisson bivector on $M$ (thus depending only of $X$) and $f_2^{ij}=-\partial_{k}\Pi^{ij} Y^{k}+R^{ij}(X)$, where $R^{ij}$ is a bivector satisfying \eqref{RPoisson} for $p=1$. The resulting theory is the one obtained from \eqref{Sp+1}, respectively \eqref{Sp+1cov} in covariant form, for $p=1$ with the associated gauge transformation, field equations and covariant formulation. Therefore, 
 \be 
 S^{(2)}_+[X,Y,A,Z]|_{f_3=0, f_1=\Pi, f_2^{ij}=R^{ij}-\partial_k\Pi^{ij}Y^{k}}=S^{(2)}[X,Y,A,Z]\,.
 \ee
 As a final simple example, we note that choosing $f_1=0$, $f_2=Y^{\ast}(\widetilde{\Pi})$ and $f_3=X^{\ast}(\Pi)$ with $H(X,Y)=X^{\ast}H+Y^{\ast}\widetilde{H}$, for two closed 3-forms $H$ and $\widetilde{H}$ and two $H-$ and $\widetilde{H}-$twisted Poisson bivectors $\Pi$ and $\widetilde{\Pi}$, one obtains the action for two (uncoupled) HPSMs, i.e.
 \be 
 S^{(2)}_+[X,Y,A,Z]|_{f_1=0}=S_{\text{HPSM}}[X,Z]-S_{\text{HPSM}}[Y,A]\,.
 \ee  
 These belong to a class of models where the target space is essentially doubled. The realization of the target space in terms of $Q$-manifolds is that of the double (degree-shifted) cotangent bundle $T^{\ast}[1]T^{\ast}[1]M$. It is worth mentioning that doubled sigma models in two dimensions have found applications in the context of string theory backgrounds with manifest T-duality, see e.g. \cite{Berman:2013eva}. We will not explore this connection further in the present paper though, since it deserves a separate analysis.    
 
 \subsubsection{Bi-twisted R-Poisson 3D TFT and twisted Courant sigma models}
 \label{sec322}
 
 Moving on to three dimensions and returning to the target space covariant formulation, we consider examples of $S^{(3)}$ as given in \eqref{Sp+1cov} for $p=2$ and deformed by the corresponding admissible terms from Table \ref{table2}. Thus we are led to the action functional 
 \bea \label{S3+}
 S^{(3)}_+=\int_{\S_{3}}\left(\langle {Z}^{\mathring{\nabla}},F\rangle-\langle Y,G^{\mathring{\nabla}}\rangle+\frac 1{3!}R(A,A,A)\right)+\int_{\S_{4}}X^{\ast}H_4 +S^{(3)}_{\text{def}}\,.
 \eea 
 Note that we have switched to the notation where the composition of all background data with the base map $X$ is implicit. Moreover, $R$ is now a 3-vector on $M$ and $\Pi$ is a 2-vector, not necessarily Poisson. In the absence of $S_{\text{def}}$ this is simply $S^{(3)}$ with its associated twisted $R$-Poisson structure of order 3, in which case $\Pi$ is Poisson. A less general version of such cases in three dimensions was considered in \cite{Chatzistavrakidis:2015vka,Bessho:2015tkk,Heller:2016abk,Chatzistavrakidis:2018ztm}. Specifically, in these references (i) only a non covariant formulation was found and thus the theories were defined only in a local patch, (ii) the Wess-Zumino term due to $H_4$ was absent and (iii) due to the absence of $H_4$, the $R$-Poisson structure was untwisted, namely $[\Pi,R]=0$. Presently, we go beyond all above restrictions, and in addition to that we eventually allow $\Pi$ to be a twisted Poisson bivector too.  
 Hence, let us examine what happens when the deformation is included. First of all, based on the discussion of Section \ref{sec4}, the most general deformation action{\footnote{To be precise, one may allow for an arbitrary function in front of the first term, however we consider this to be the identity here.}} is 
 \bea \label{Sdef3}
 S_{\text{def}}^{(3)}=\int_{\S_3}\left(\langle{Z}^{\mathring{\nabla}},Y\rangle +\frac 12 f(Y,Y,A)+\frac 1{3!}S(Y,Y,Y)\right)\,,
 \eea    
 where $f\in \G(TM\otimes \bigwedge^{2}T^{\ast}M)$ and $S\in \G(\bigwedge^3T^{\ast}M)$. That these deformations exist in three dimensions should not come as a surprise. Indeed, the most general AKSZ topological field theory in three dimensions is the Courant sigma model and \eqref{S3+} is of this type for $H_4=0$. This may be seen by realizing that the fields $A$ and $Y$ are both spacetime 1-forms for $p=2$ and therefore they may be combined into 1-field ${\cal A}\in \Omega^{1}(\S_3,X^{\ast}(TM\oplus T^{\ast}M))$. Then the full action may be written in the form 
 \be 
S^{(3)}_+= \int_{\S_{3}}\left(\langle  Z^{\mathring{\nabla}},\dd X+Y\rangle-\langle A,{\cal F}^{\mathring{\nabla}}\rangle+ \frac{1}{3!}\widehat{T}({\cal A},{\cal A},{\cal A}) \right)+\int_{\S_{4}}X^{\ast}H_4\,,
 \ee    
 where we made use of the second expression in \eqref{Sp+1cov} and the tensor $\widehat{T}$ is an element of  $\G(\otimes^{3}(TM\oplus T^{\ast}M))$.   
 We note that this target space covariant formulation of the Courant sigma model does not appear in the original publications \cite{Ikeda:2002wh,Hofman:2002jz,Roytenberg:2006qz}, this being another side result of our approach. What is more, we have introduced an additional four-dimensional Wess-Zumino term in the theory. A similar term was considered in a different context in \cite{Hansen:2009zd}.
 
 Let us now examine this model in more detail. First, we note that for vanishing $\Pi, f, R$ and $H_4$ one obtains directly the sigma model corresponding{\footnote{This correspondence is established in \cite{Roytenberg:2006qz} and we will not revisit it here.}} to the \emph{standard} Courant algebroid with 3-form twist, namely to the quadruple $(E, [\cdot,\cdot]_{E},\langle\cdot,\cdot\rangle,\rho_0:E\to TM)$ with $E$ being the extended vector bundle $TM\oplus T^{\ast}M$, the bracket on sections $v\oplus \eta, v'\oplus\eta',\in \G(E)$ with $v, v'\in \G(TM)$ and $\eta, \eta'\in \G(T^{\ast}M)$ being 
 \be 
 [v\oplus\eta,v'\oplus \eta']_{E}=[v, v']\oplus \left({\cal L}_{v}\eta'-\iota_{v'}\dd\eta-\iota_{v}\iota_{v'}S\right)\,,
 \ee 
 the canonical non-degenerate inner product being 
 \be 
 \langle v\oplus \eta,v'\oplus \eta'\rangle =\frac 12 (\iota_{v}\eta'+\iota_{v'}\eta)\,,
 \ee 
 and the anchor map $\rho:TM\oplus T^{\ast}M\to TM$ given as 
 \be 
 \rho_0=\text{id}\oplus 0\,.
 \ee 
 This is certainly the simplest possibility, however it is not the only one and in particular it is rather orthogonal to the spirit of the present paper. Here we are more interested in examples where the assumptions of vanishing $\Pi, f, R$ and $H_4$ are relaxed and we would like to determine the underlying structure on the target space in terms of suitably (bi-)twisted $R$-Poisson structures. In particular, we employ the following definition 
 \begin{defn}\label{bitRP}
 	A bi-twisted $R$-Poisson manifold is a quintuple $(M,\Pi,R,S,H)$ consisting of a smooth manifold $M$ equipped with an antisymmetric bivector $\Pi$, an antisymmetric 3-vector $R$, a $3$-form $S$ 
 and a $4$-form $H$ such that 
 \bea \label{bitRpoisson}
 \frac 12 \, [\Pi,\Pi]&=&R+\langle \Pi\otimes \Pi\otimes \Pi,S\rangle\,,\\[4pt] 
 \dd S&=&-H\,.
 \eea 
By the second condition, $H$ is not only closed but also exact.
 \end{defn}
We observe that for $R=0=H$, this definition reduces to the one of a $S$-twisted Poisson manifold with a closed 3-form $S$. In general though, $R$ and $H$ are not zero and therefore $S$ is not closed. In fact, \eqref{bitRpoisson} may be viewed as a definition of $R$ in terms of $\Pi$ and $S$; when it is zero, the structure is twisted Poisson. Moreover, it is useful to note that due to the fact that the Schouten-Nijenhuis bracket satisfies the Jacobi identity 
\be 
[[u,v],w]=[u,[v,w]]-(-1)^{(|u|-1)(|v|-1)}[v,[u,w]]\,,
\ee 
where $|u|$ denotes the degree of the multivector field $u$, it trivially follows that 
\be 
[\Pi,[\Pi,\Pi]]=0\,,
\ee 
for an arbitrary bivector $\Pi$. In the standard twisted Poisson case, this has the immediate consequence that 
\be 
[\Pi,\langle \Pi\otimes\Pi\otimes\Pi,S\rangle]=0\,,
\ee 
which is in fact equivalent to the closure of the 3-form $S$. In the bi-twisted $R$-Poisson case though, this ceases to be true. Instead the 3-form $S$ is not closed and taking the Schouten-Nijenhuis bracket of Eq. \eqref{bitRpoisson} with $\Pi$ leads to the identity 
\be\label{piR}
 [\Pi,R]= \langle \Pi\otimes\Pi\otimes\Pi\otimes\Pi,H\rangle+\frac{1}{4} \, \mathrm{Alt}\, \langle\Pi\otimes\Pi\otimes R,S\rangle\,, 
\ee 
 where $\mathrm{Alt}$ denotes antisymmetrization in the fourfold tensor product of $T^{\ast}M$. This identity is useful in establishing gauge invariance of the corresponding topological field theory that we discuss below. 
 
  These structures are associated to \emph{non-standard} twisted Courant algebroids, already described in \cite{liu}, namely choosing a more general anchor 
 \be \label{rhogeneral}
 \rho=(\l\, \text{id})\oplus \Pi\,,
 \ee  
 with $\Pi$ an antisymmetric bivector. We introduced the real parameter $\l$ to account for two separate cases obtained for $\l=0$ and $\l=1$. The former case essentially corresponds to switching off the first term in the deformation action \eqref{Sdef3}. We will mostly be interested in the latter case though, since this can give rise to more general structures than the twisted $R$-Poisson one that has already been covered in detail earlier. Therefore, we set $\l=1$ from now on. Note that for non-standard Courant algebroids, the generalization of the bracket was also already found in \cite{liu} (in its antisymmetrized form) to be 
 \be 
  [v\oplus\eta,v'\oplus \eta']_{E}=\left([v, v']+{\cal L}_{\eta}v'-\iota_{\eta'}v\right)\oplus \left([\eta,\eta']+{\cal L}_{v}\eta'-\iota_{v'}\dd\eta-\iota_{v}\iota_{v'}S\right)\,,
 \ee    
 where $[\eta,\eta']$ denotes a suitable bracket of 1-forms, typically the Koszul-Schouten bracket. 
 
 We now momentarily switch to the local coordinate formulation for clarity. Consider the action functional 
 \begin{align} 
 S_{+}^{(3)}&=\int_{\S_3}\left(Z_i\w(\dd X^{i}+\Pi^{ij}A_j+Y^{i})-A_i\w\dd Y^{i}-\frac 12  \left(\partial_k\Pi^{ij}-\Pi^{il}\Pi^{jm}S_{lmk}\right)Y^{k}\w A_i\w A_j\right. \nn\\[4pt]
 &\left.-\frac 12 \Pi^{kl}S_{ijl}Y^{i}\w Y^{j}\w A_k+\frac 1{3!}R^{ijk}A_i\w A_j\w A_k+\frac 1{3!}S_{ijk}Y^{i}\w Y^{j}\w Y^{k}\right)+\int_{\S_4}X^{\ast}H_4\,. \nn\\ \label{S3++}
 \end{align} 
 Note that the background data depend precisely on the components of $\Pi, R, S$ and $H$. Next we introduce the following set of gauge transformations (neglecting wedge products between differential forms,)
 \bea
 \d^{+}X^{i}&=&\d X^{i}+\chi^{i}\,, \label{gt1+}\\[4pt]
\label{gt2+} \d^{+}A_{i}&=&\d A_i+\psi_i+S_{ijk}Y^{j}\chi^{k}+\Pi^{kl}S_{lij}(Y^{j}\epsilon_k+\chi^{j}A_k)+\Pi^{jl}\Pi^{km}S_{mli}A_j\epsilon_k\,,\\[4pt]
\label{gt3+} \d^{+}Y^{i}&=&\d Y^{i}+\Pi^{il}\Pi^{km}S_{lmj}(\chi^{j}A_k+Y^{j}\epsilon_k)+\Pi^{il}S_{ljk}\chi^{j}Y^k\,,\\[4pt] 
 \d^{+}Z_{i}&=& \d Z_i|_{\Omega\to 0} +\partial_i(\Pi^{km}\Pi^{ln}S_{jmn})(Y^{j}A_k\epsilon_{l}-\frac 12 A_kA_l\chi^{j}) -\frac 12 \partial_i S_{jkl}Y^{j}Y^{k}\chi^{l}- \nn\\[4pt] && -\,\partial_i(\Pi^{jm}S_{mkl})(Y^{k}A_j\chi^{l}+\frac 12 Y^{k}Y^{l}\epsilon_j)+\frac 1{3!}H_{ijkl}(\Pi^{mj}\epsilon_m+\chi^{j}){\Omega}_{+}^{kl}\,,\label{gt4+}
 \eea 
 where $\d(\dots)$ are the transformations of the undeformed theory given in \eqref{gt1}-\eqref{gt4} for $p=2$ and the quantity ${\Omega}_{+}^{kl}$ is given by 
 \be \label{Omega+}
 {\Omega}_{+}^{kl}=\dd X^{k}\w\dd X^{l}-\dd X^{k}\w (Y^{l}+\Pi^{lm}A_m)+(Y^{k}+\Pi^{km}A_m)\w (Y^{l}+\Pi^{ln}A_n)\,,
 \ee 
 essentially generalizing the $\Omega^{kl}$ that appeared in the undeformed case. A calculation along the same lines as described in Section \ref{sec22} establishes the following statement, once one takes into account Definition \ref{bitRP} and the identity \eqref{piR}. 
  \begin{prop}\label{propS3+}
  	The classical action functional $S^{(3)}_{+}$ given in \eqref{S3++} on a 3-dimensional spacetime $\S_{3}$ without boundary with Wess-Zumino term supported on a four-dimensional manifold $\S_4$ such that $\partial\S_4=\S_3$ is invariant under the gauge transformations \eqref{gt1+}-\eqref{gt4+} if and only if the target space M is a bi-twisted R-Poisson manifold and ${\Omega}_{+}^{kl}$ is given as in Eq. \eqref{Omega+}.  
  \end{prop} 
 \begin{rmk}
 	A special case is obtained if we set $R=0$ and $H_4=0$. Then $S$ is a closed 3-form and $\Pi$ is an $S$-twisted Poisson structure. This is then a 3-dimensional topological field theory with twisted Poisson structure. It shares the same structure as the 2-dimensional HPSM but on the other hand they differ in that there is now no Wess-Zumino term supported in a higher-dimensional bulk.  
 \end{rmk}

\begin{rmk}
	To avoid confusion, we note that absence of $S$ does not imply that $H=0$. This is clear from the fact that we have already constructed $S^{(p+1)}$ where this is the case, but it is somewhat obscured by definition \ref{bitRP}, where the last equation seems to imply it. This is however not true, since in the spirit of \eqref{rhogeneral} we could think of the deformation action accompanied by the real parameter $\l$, which we set to one above, namely $S^{(3)}_+=S^{(3)}+\l S_{\text{def}}$. Then $\l=0$ results in a twisted $R$-Poisson structure as in Proposition \ref{propSp+1} for $p=2$, whereas $\l=1$ results in a bi-twisted $R$-Poisson structure as in Proposition \ref{propS3+}. This is how twisted and bi-twisted $R$-Poisson structures and their associated topological field theories are unified in a single picture.  
\end{rmk}
 
 \subsubsection{The covariant form of the bi-twisted R-Poisson TFT and its Q-structure}
 \label{sec333}
 
 To complete the discussion on the topological field theory induced by a bi-twisted $R$-Poisson structure, we now discuss its target space covariant formulation. Unlike the undeformed model, which corresponds to $S=0$ and its covariance is attained by means of the connection $\mathring{\nabla}$ without torsion, the case we discuss here requires a new connection to be covariantized. In particular, similarly to the HPSM, we introduce a torsional connection $\nabla$ on $TM$ with connection coefficients{\footnote{Note that in our conventions for the three-dimensional theory this connection differs by a sign in its torsional part in comparison to the one used in \cite{Ikeda:2019czt} and mentioned in Section \ref{sec21}.}} 
 \be \label{gamma}
 \G^{i}_{jk}=\mathring{\G}^{i}_{jk}+\frac 12 \Pi^{il}S_{ljk}\,.
 \ee   
 This connection acts in an obvious way on a coordinate basis of $TM$ and induces one on $T^{\ast}M$ in a straightforward way. The same holds for the corresponding induced covariant differential on forms, which we now denote by $\DD$. The coefficients $\G^{i}_{jk}$ are no longer symmetric in their lower indices and the antisymmetric part is precisely the torsion of $\nabla$,
 \be 
 \Theta^{i}_{jk}=\G^{i}_{jk}-\G^{i}_{kj}=\Pi^{il}S_{ljk}\,.
 \ee   
 In geometric terms, this expression may be written as 
 \be\label{Theta} 
 \Theta=\langle\Pi,S\rangle\,,
 \ee 
 with $\Theta$ being as usual a vector-valued 2-form, namely a section of $TM\otimes \Omega^{2}(M)$. 
 
 Equipped with the above connection, we can directly covariantize all expressions of Section \ref{sec322}, namely the action functional, the field equations obtained from it and the gauge transformations. The approach is the same as the one we followed in the general case in section \ref{sec22} and therefore we will be somewhat brief here, highlighting the differences due to the torsion. 
 We focus on the field equations of the action \eqref{S3++}, denoting the corresponding field strengths by a hat to avoid confusion with the undeformed theory. They read as 
 \bea 
 \widehat{F}^{i}&=&F^{i}+Y^{i}=0\,,\\[4pt]
 \widehat{G}_{i}&=&G_{i}-Z_i-\frac 12 \Pi^{kl}\Pi^{jm}S_{lmi}A_k\w A_j+\Pi^{kl}S_{ijl}Y^{j}\w A_k-\frac 12 S_{ijk}Y^{j}\w Y^{k}=0\,, \\[4pt]
\label{eom3+} \widehat{\cal F}^{i}&=&{\cal F}^{i}-\Pi^{il}\Pi^{jm}S_{lmk}A_j\w Y^{k}+\frac 12 \Pi^{il}S_{jkl}Y^{j}\w Y^{k}=0\,, \\[4pt]
\widehat{\cal G}_{i}&=&{\cal G}_{i}+\frac 12 \partial_i(\Pi^{jl}\Pi^{km}S_{lmn})Y^{n}\w A_j\w A_k-\nn\\[4pt] && \quad \,-\,\frac 12 \partial_{i}(\Pi^{kl}S_{jml})Y^{j}\w Y^{m}\w A_k+\frac 1 {3!}\partial_iS_{jkl}Y^{j}\w Y^{k}\w Y^{l}\,, 
 \eea 
 where the unhatted ones are given in \eqref{eom1}-\eqref{eom4} for $p=2$. The first equation is simple, since it is already covariant, and it may be written as 
 \be 
 \widehat{F}=F+Y\,.
 \ee  The second one contains no partial derivatives and therefore its covariantization is also direct. Specifically, recalling the form of $G_{i}$ from Eq. \eqref{Ginter}, it is just a matter of using the new connection $\nabla$. Using that 
 $\DD A_i=\dd A_i-\Gamma^{k}_{ij}\dd X^{j}\w A^{k}$ and redefining the field $Z$ to ${Z}_{i}^{\nabla}=Z_i+\G^{k}_{ij}Y^{j}\w A_k$, one finds 
 \be 
 \widehat{G}_{i}=\DD A_i+\frac 12\mathring{\nabla}_i\Pi^{jk}A_j\w A_k-{Z}^{\nabla}_i-\G^{k}_{ij}A_k\w\widehat{F}^{j}-\Pi^{kl}S_{iml}A_k\w Y^{m}-\frac 12 S_{ijk}Y^{j}\w Y^{k}\,.
 \ee 
 Recalling the definition \eqref{Theta}, that $T=-\mathring{\nabla}\Pi$ (which does not change due to the torsion of the connection $\nabla$) and defining the covariant field strength $\widehat{G}^{\nabla}_i=G_i+\G^{k}_{ij}A_k\w \widehat{F}^{j}$, the final fully covariant result is 
 \be 
 \widehat{G}^{\nabla}=\DD A-{Z}^{\nabla}-\frac 12 T(A,A)-\Theta(A,Y)-\frac 12 S(Y,Y)\,.
 \ee 
 We emphasize that this (and the following) covariant expression is not unique, a typical feature of torsional geometries. Indeed, torsion terms can be absorbed in the definition of connection coefficients and in particular note that $\G^{i}_{jk}$ of \eqref{gamma} and the alternative connection with coefficients $\mathring{\Gamma}^{i}_{jk}-\frac 12 \Pi^{il}S_{ljk}$ differ precisely by the torsion. 
  
 Next we turn to the field strength of the field $Y$, and we again observe that there are no additional partial derivatives in the field equation \eqref{eom3+}. A similar manipulation in this case indicates the definition $(\widehat{\cal F}^{\nabla})^{i}=\widehat{\cal F}^{i}-\G^{i}_{jk}Y^{k}\w\widehat{F}^{j}$, in which case the covariant tensor is 
 \be 
 \widehat{\cal F}^{\nabla}=\DD Y+\Pi({Z}^{\nabla})-T(A,Y)+\frac 12 R(A,A)+\Theta(Y,Y)-2\langle \Pi,\Theta(Y,A)\rangle\,.
 \ee  
 The covariantization of the final field strength is slightly more demanding due to the appearance of new terms with partial derivatives. 
 It is a straightforward exercise to show that 
 \bea 
 \widehat{\cal{G}}_{i}&=&-\DD Z^{\nabla}_i+(T_i^{km}+\Pi^{lk}\Theta^{m}_{il})Z^{\nabla}_m\w A_k+\frac{1}{3!}(\nabla_iR^{jkl}+{\cal T}_{i}^{jkl})A_j\w A_k\w A_l+ \nn\\[4pt] 
 \quad && + \, \frac 1{3!} (\nabla_iS_{jkl}+H_{ijkl})Y^{j}\w Y^k\w Y^l-\frac 12({\cal R}^{l}_{ijk}+{\nabla}_i\Theta^{l}_{jk}-\Pi^{ml}H_{ijkm})Y^j\w Y^k\w A_l + \nn\\[4pt] 
 \quad && +\, \left(\frac 12 \nabla_iT_q^{kn}-\Pi^{ln}({\cal R}^k_{qli}+\mathring{\nabla}_l\Theta^{k}_{iq}-\frac 12\Theta^{m}_{iq}\Theta^{k}_{lm}+\Theta^{k}_{im}\Theta^{m}_{lq}-\frac 12 \Pi^{mk}H_{iqml})+\Theta^{n}_{lq}T_i^{kl}\right)\times \nn\\[4pt]  
 \quad && \times \, A_k\w A_n\w Y^q + \Gamma^{k}_{ij}A_k\w \widehat{\cal F}^{j}-\G^{j}_{ik}Y^k\w\widehat{G}_{j}+U_{ij}\w\widehat{F}^{j}\,,
 \eea 
 where 
 \be 
 U_{ij}=-\G^{k}_{ij}Z^{\nabla}_{k}+(\partial_j\G^{k}_{il}-\frac 12 \Pi^{nk}H_{ilnj})Y^l\w A_k+\frac 13 H_{ilmj}Y^m\w Y^l-\frac 1{3!}H_{ijlm}\Omega^{lm}+\frac 1{3!}H_{iljm}\widehat{F}^m\w Y^l
 \ee 
 is an auxiliary 2-form that we will not attempt to simplify since it does not play any further role in the analysis. 
 Here we have defined the curvature ${\cal R}$ of the connection $\nabla$, which in components and with respect to the curvature of the torsionless one $\mathring{\nabla}$ and the torsion $\Theta$  takes the form 
 \bea 
 {\cal R}^{k}_{ipq}=\mathring{\cal R}^{k}_{ipq}+\mathring{\nabla}_{[p}\Theta^{k}_{\underline{i}q]}+\frac 12 \Theta^{k}_{m[p}\Theta^{m}_{\underline{i}q]}\,.
 \eea 
 Moreover, ${\cal T}\in\G(T^{\ast}M\otimes \bigwedge^{3}TM)$ is in the present case given in components by
 \be 
 {\cal T}_i^{jkl}=\Pi^{jm}\Pi^{kn}\Pi^{lp}H_{mnpi}\,.
 \ee  
 Note also that the covariant derivative on $R$ with respect to the connection with torsion and similarly the one on $\Theta$ itself are given as 
 \bea 
 \nabla_{i}R^{jkl}&=&\mathring{\nabla}_iR^{jkl}+\frac 32 \Theta^{j}_{im}R^{klm}\,, \\ 
  \nabla_i\Theta^{k}_{pq}&=&\mathring{\nabla}_i\Theta^{k}_{pq}+\frac 12 \Theta^k_{il}\Theta^{l}_{pq}-\Theta^{l}_{i[p}\Theta^{k}_{\underline{l}q]}\,. 
 \eea 
 The target space covariant field strength is then defined through $\widehat{\cal G}^{\nabla}_i:=\widehat{\cal G}_i-\G_{ij}^kA_k\w\widehat{\cal F}^j+\G^j_{ik}Y^k\w\widehat{G}_{j}-U_{ij}\w \widehat{F}^j$, which is now given only in terms of tensorial quantities and gives rise to the field equation $\widehat{\cal G}^{\nabla}=0$.
 
 Aside the field equations, a similar approach gives the covariant form of the gauge transformations, which we do not record here. Finally, the classical action functional in covariant form reads as 
 \bea 
 S^{(3)}_{+}&=&\int_{\S_3}\left(\langle Z^{\nabla}-\frac 12 \Theta(Y,A),\widehat{F}\rangle-\langle Y,G^{\mathring{\nabla}}\rangle+\frac 1{3!}R(A,A,A)+\frac 1{3!}S(Y,Y,Y)-\right. \nn\\[4pt] \\ && \qquad  \left. -\, \frac 12\Theta(Y,Y,A)   - \frac 12 \text{Alt}(\iota_{\rho}\Theta)(Y,A,A)\right)+\int_{\S_4}X^{\ast}H_4\,.  
 \eea 
Certainly there are more than one ways to express this action functional in such a target space covariant form by reshuffling torsion terms. 
 
 To conclude this example, we present the associated $Q$-manifold. The homological vector field in this case becomes 
 \bea 
 Q_+&=&Q-y^{i}\partial_{x^{i}}+\left(Z_{i}+\frac 12 \Pi^{jm}\Theta^{k}_{mi}a_ka_j-\Theta^{k}_{ij}y^ja_k+\frac 12 S_{ijk}y^jy^k\right)\partial_{a_i}+\nn\\[4pt]
 && + \left(\Pi^{jm}\Theta^{i}_{mk}a_jy^{k}-\frac 12 \Theta^{i}_{jk}y^jy^k\right)\partial_{y^{i}}- \nn\\[4pt] &&-\left(\frac 12 \partial_{i}(\Pi^{km}\Theta^{j}_{mn})y^na_ja_k+\frac 12 \partial_i\Theta^{k}_{ml}y^{j}y^ma_k-\frac 1{3!}\partial_iS_{jkl}y^jy^ky^l\right)\partial_{z_i}\,,
 \eea
 where $Q$ is given as in \eqref{Qp+1} for $p=2$. The correspondence now is that $(T^{\ast}[2]T^{\ast}[1]M,Q_+)$ is a $Q$-manifold if and only if $(M,\Pi,R,S,H)$ is a bi-twisted $R$-Poisson manifold. 
 
 \subsubsection{Twisted Tetravector-Poisson 4D TFT}
\label{sec334}

 As a final example, let us consider the four-dimensional case, which is obtained for $p=3$. Such examples were considered first in \cite{Ikeda:2010vz} from the point of view of $QP$ structures of degree 3 and homotopy algebroids. Employing the AKSZ construction, the authors constructed examples of four-dimensional topological field theories. In a similar spirit, yet another example of this kind was found in \cite{Chatzistavrakidis:2019seu} with an underlying Poisson structure and a 4-vector $R$ satisfying $[\Pi,R]=0$. Here we are interested in cases that go beyond both above approaches, in that we do not have a vanilla $QP$ structure due to the twist by a 5-form $H$, and for the same reason $[\Pi,R]\ne 0$. As a byproduct of the example presented below, in absence of the 5-form we also find the covariant version of the model presented in \cite{Chatzistavrakidis:2019seu}, which was only defined in a local coordinate patch.
 
The example is a simple application of what we have constructed in sections \ref{sec2} and \ref{sec3}. The target space graded manifold is $T^{\ast}[3]T^{\ast}[2]M$ and the fields of the theory are the scalars, $X^{\ast}T^{\ast}M$-valued 1-forms, $X^{\ast}TM$-valued 2-forms and $X^{\ast}T^{\ast}M$-valued 3-forms. The background fields are the Poisson structure $\Pi$, a tetravector $R$ and a 5-form $H$, giving in a local coordinate patch the action functional 
\bea 
S^{(4)}&=&\int_{\S_{4}}\left(Z_i\w\dd X^{i}-A_i\w \dd Y^{i}+\Pi^{ij}(X)\, Z_i\w A_j-\frac 12 \, \partial_k\Pi^{ij}\, Y^{k}\w A_i\w A_j\, + \right. \nn \\[4pt] &&  \qquad \left. +\, \frac 1{4!}R^{ijkl}(X)\, A_{i}\w A_j\w A_k \wedge A_{l}\right)+\int_{\S_{5}}\frac 1{5!}H_{ijklm}(X)\dd X^{ijklm}\,,\label{S4}
\eea     
where $\dd X^{ijklm}\equiv \dd X^{i}\w\dd X^{j}\w\dd X^{k}\w\dd X^{l}\w\dd X^{m}$. For $H=0$ this is precisely the theory considered in section 4.3 of \cite{Chatzistavrakidis:2019seu}. From Proposition \ref{propSp+1} we know that this theory is induced by a twisted $R$-Poisson structure of order 4, also for $H\ne 0$. This means that 
\be 
[\Pi,R]=-\langle\otimes^5\Pi,H\rangle\,,
\ee  
or in local coordinates
\be 
5\Pi^{[j\underline{i}}\partial_iR^{klmn]}+10R^{[j\underline{i}kl}\partial_i\Pi^{mn]}=-\Pi^{jj'}\Pi^{kk'}\Pi^{ll'}\Pi^{mm'}\Pi^{nn'}H_{j'k'l'm'n'}\,.
\ee
What is more, apart from the generalisation due to the inclusion of the Wess-Zumino term, our analysis in Section \ref{sec3} has now produced the target space covariant form of the action \eqref{S4}. Indeed, this is simply given by \eqref{Sp+1cov} or \eqref{Sp+1covb} for $p=3$. 

Finally, recall that four is the maximum dimension of $\S$ such that a deformation is admissible with the given field content. Presently there exists only a single possible deformation term, which is 
\be \label{S4+}
S_{\text{def}}^{(4)}=\int_{\S_{4}}\frac 12\, g_{ij}(X) Y^{i}\w Y^{j}\,.
\ee 
We denoted the background field by $g_{ij}$ since it is a symmetric tensor, reminiscent of a metric. This is an interesting term in itself. For a Lie-algebra valued 2-form field strength in four dimensions $F=(F^{a})$, the index $a$ being a Lie algebra one, there exists a theta term 
\be 
S_{\theta}= \frac{\theta}{32\pi^{2}}\,\int  \d_{ab} \, F^{a}\w F^{b}\,.
\ee 
This is the second Lorentz and gauge invariant quadratic term in four dimensions aside the kinetic term and it is the backbone of topological Yang-Mills theory and also the QCD CP-violating theta term. One should note that such a term also exists in the Abelian theory, with certain physical applications such as in the effective field theory of topological insulators \cite{Qi:2008ew} and topological superconductors \cite{Qi:2012cs}.   
The deformation \eqref{S4+} is of this nature, at least if one thinks of $Y^{i}$ as an exact $(p-1)$-form, and it corresponds to a generalised theta term in presence of multiple fields, and therefore it is not surprising that it comes with a symmetric background field (see also \cite{Chatzistavrakidis:2020kpx} for further explanations on this point.)  However, in the spirit of the present paper we would like to think of this term as accompanying the theory given by the action functional \eqref{S4}, and also of $Y^{i}$ as a gauge field rather than a field strength. In such a case, one should determine the modified transformation rules and the underlying structure that yields this functional gauge invariant. For $H=0$, this is a special case of \cite{Ikeda:2010vz} and one can quickly see that a necessary condition is 
\be 
\Pi^{ij}g_{jk}=0\,.
\ee 
This does not change in presence of $H$. This condition is rather strong and it only allows degenerate cases. Therefore we will not examine such deformed models in four dimensions further.

\section{Discussion and Conclusions} 
\label{sec6}

We presented a large class of topological field theories with Wess-Zumino term, induced by a twisted $R$-Poisson structure and extensions thereof. These models can be thought as generalizations of the two-dimensional Poisson and twisted Poisson sigma models in any dimension. Indeed, a twisted $R$-Poisson structure includes a Poisson 2-vector accompanied by a $(p+1)$-vector $R$ and a closed $(p+2)$-form that satisfy the structural condition \eqref{RPoissonintro}. The topological field theories we constructed are gauge theories whose gauge structure is compatible with a target space being a twisted $R$-Poisson manifold. We determined the action functional, gauge transformations and field equations of these theories in a local patch and then we carefully extended them in a target space covariant formulation by means of an auxiliary connection. In addition, we showed that the structure of the target space as a differential graded manifold is $T^{\ast}[p]T^{\ast}[1]M$ and determined the corresponding homological vector field and symplectic structure.

 Studying possible deformations of the general case, we classified the admissible ones which exist only in dimensions 2, 3 and 4 under the assumed field content. Subsequently, we investigated islands of theories in these special dimensions that are induced by the deformations. Notably, we found a three-dimensional theory induced by a twisted Poisson structure that extends the two-dimensional one. The three-dimensional models with twisted Poisson and twisted $R$-Poisson structure can be elegantly combined in a more general setting, controlled by a bi-twisted $R$-Poisson structure that includes the other two as special cases. The latter contains an additional non-closed 3-form $S$ aside the 2- and 3-vectors and 4-form of the $R$-Poisson case, and is defined through the structural condition \eqref{bitRpoissonintro}, where the 2-vector departs from being Poisson in a general way, its Schouten-Nijenhuis bracket with itself receiving contributions from both the 3-vector $R$ and the 3-form $S$. For this case, we also presented the target space covariant formulation, this time using an auxiliary connection with torsion, and the corresponding differential graded supermanifold picture. 
 
 One of the basic questions that arises regards the BV action of the topological field theories we presented here. For the case that the Wess-Zumino term is absent this can be done using the AKSZ construction. However, the situation is more complicated in presence of the $(p+2)$-form $H$. Already in the two-dimensional case of the twisted Poisson sigma model, the authors of \cite{Ikeda:2019czt} demonstrated that a naive generalization of AKSZ does not yield the correct BV action. The underlying reason is that the Q and P structures in the twisted case are not compatible and therefore the target graded manifold is not QP. Instead they followed a direct approach and they found the BV action constructively. This amounts to defining the BRST operator and introducing a ghost for the single scalar gauge parameter of the theory. Recognizing that the BRST operator on the 1-form field does not square to zero but instead it is proportional to the field equation for it, necessitates the use of antifields. A careful analysis then shows that the BV action contains three contributions with 0, 1 and 2 antifields respectively. Notably, the contribution with 2 antifields turns out very simple, consisting of a single term with the same coefficient as the one appearing in the action of the square of the BRST operator on the 1-form. Yet, it is precisely due to this term that the classical master equation for the model is satisfied, or alternatively, that the classical master equation for the naive generalization of AKSZ is violated when the twist in non-vanishing (see Appendix B of \cite{Ikeda:2019czt}).  
 
 The above statements persist in the theories we constructed here too and therefore one should follow the standard field antifield formalism to determine the BV action. However, the situation is significantly more complicated than the two-dimensional case. This comes about because the gauge parameters are now higher degree differential forms. Consequently, unlike the twisted Poisson sigma model, one must introduce a tower of ghosts for ghosts both for the ghost from the $(p-2)$-form gauge parameter $\chi$ and for the one from the $(p-1)$-form gauge parameter $\psi$. As usual, this is simply due to the fact that the gauge parameters are not independent, since there exist classes of parameters for which the gauge transformations vanish. In other words, the twisted Poisson sigma model is a theory with a gauge algebra that closes only on-shell yet it is an irreducible theory, whereas the cases we studied apart from having an open gauge algebra they also correspond to reducible theories of many stages, as typical for theories that contain fields of form degree higher than 1 \cite{HT}. In relation to this, one can notice that the BRST operator does not square to zero not only for a single field of the theory but for several of them (all of $A, Y$ and $Z$) and what is more the same holds for some of the ghosts and all ghosts for ghosts (unlike the twisted Poisson sigma model where the action on the single ghost is zero and there are no ghosts for ghosts). This by no means implies that in our case the situation is intractable, on the contrary this direct approach is expected to work. However, because of the complexity and subtle points we mentioned here, it requires a separate treatment and we plan to report on this in a forthcoming publication \cite{prep}.      
 
 Besides the BV action, it would be interesting to investigate the relation of the (bi-)twisted $R$-Poisson topological field theories to constructions in terms of $L_{\infty}$ algebras, for example within the higher gauge theory approach of \cite{Gruetzmann:2014ica}. Interestingly, a direct construction of membrane sigma models in terms of $L_{\infty}$ algebras was recently discussed in \cite{Grewcoe:2020ren} and also extended to \emph{curved} $L_{\infty}$ algebras \cite{Grewcoe:2020gka}. From a different point of view, it would also be desirable to understand twisted $R$-Poisson structures in the context of $P_{\infty}$ (homotopy Poisson) structures described in \cite{Voronov} (see also \cite{Kupriyanov:2021cws} for recent applications of this idea).  
 
 Finally, one should keep in mind that in this paper we fixed the field content of the theories from the beginning, essentially motivated by the structures we were aiming at utilizing and the specific questions we posed in the introduction. Other field contents can yield interesting and useful topological field theories too with target space graded supermanifolds other than $T^{\ast}[p]T^{\ast}[1]M$. For instance one could consider cases with target space being the differential graded symplectic manifold $T^{\ast}[p]\left(\bigwedge^{p-1}\right)[p-1]T[1]M$ considered in \cite{Bouwknegt:2011vn} in the context of Nambu structures and AKSZ constructions for $p$-branes. Note, however, that \cite{Bouwknegt:2011vn} shows that Nambu structures cannot be twisted in the same sense as Poisson, due to the additional algebraic condition satisfied by a Nambu structure aside the differential one. Nevertheless, such approaches in general dimensions, including the one we employed in this paper, can be useful in the study of branes. String and M theory contain a variety of them, often with unconventional properties and geometry, for example the so-called exotic states listed in the review \cite{deBoer:2012ma}, which can couple to potentials of high vector degree, typically mixed with form degrees as well \cite{West:2004kb}, generating corresponding Wess-Zumino terms \cite{Bergshoeff:2010xc,Chatzistavrakidis:2013jqa}.  Higher (yet ordinary) brane Wess-Zumino terms were also studied from the differential graded symplectic manifold viewpoint in \cite{Arvanitakis:2018cyo,Arvanitakis:2021wkt}.  

\paragraph{Acknowledgements.} Useful discussions with Larisa Jonke, Zolt\'an K\"ok\'enyesi and Grgur \v{S}imuni\'c are acknowledged. The author is particularly thankful to L. Jonke for substantial comments on the manuscript. This work is
supported by the Croatian Science Foundation Project ``New Geometries for Gravity and
Spacetime" (IP-2018-01-7615).

\end{document}